\documentclass[onecolumn,floats,floatfix,prd,superscriptaddress,nofootinbib]{revtex4-2}

\usepackage{graphicx,epsfig}
\usepackage{amssymb,amsmath,amsthm,amsfonts}
\usepackage{bm}
\usepackage[inline]{enumitem}
\usepackage{tensor}
\usepackage[linktocpage]{hyperref}
\usepackage[caption=false]{subfig}
\usepackage[usenames,dvipsnames]{xcolor}
\usepackage{url}
\usepackage[normalem]{ulem}
\usepackage[inline]{enumitem}
\usepackage{xspace}
\usepackage{comment}
\usepackage{cancel}
\usepackage{multirow}
\usepackage{xcolor}

\def\I{i\,}
\def\cW{{\cal W}}

\newcommand{\appropto}{\mathrel{\vcenter{
  \offinterlineskip\halign{\hfil$##$\cr
    \propto\cr\noalign{\kern2pt}\sim\cr\noalign{\kern-2pt}}}}}

\def\Julia{\textsc{Julia}\,}

\makeatletter
\def\l@subsubsection#1#2{}
\def\l@acknowledgements#1#2{}
\makeatother

\def\newacronym#1#2#3{\gdef#1{\gdef#1{#2\xspace}#3 (#2)\xspace}}
\newacronym{\bh}{BH}{black hole}
\def\bs#1{black string#1 (BS#1)\gdef\bs{BS}}
\def\bc#1{boundary condition#1 (BC#1)\gdef\bc{BC}}

\newcommand{\sapienza}{Dipartimento di Fisica, Sapienza Università 
	di Roma, Piazzale Aldo Moro 5, 00185, Roma, Italy}
\newcommand{\infn}{INFN, Sezione di Roma, Piazzale Aldo Moro 2, 00185, Roma, Italy}

\begin{document}
\title{The Great Impersonation: $\mathcal{W}$-Solitons as Prototypical Black Hole Microstates}

\author{Alexandru Dima}
\email{alexandru.dima@uniroma1.it}
\affiliation{\sapienza}
\affiliation{\infn}

\author{Pierre Heidmann}
\email{heidmann.5@osu.edu}
\affiliation{Department of Physics and Center for Cosmology and AstroParticle Physics (CCAPP), The Ohio State University, Columbus, OH 43210, USA}

\author{Marco Melis}
\email{marco.melis@uniroma1.it}
\affiliation{\sapienza}
\affiliation{\infn}

\author{Paolo Pani}
\email{paolo.pani@uniroma1.it}
\affiliation{\sapienza}
\affiliation{\infn}

\author{Gela Patashuri}
\email{patashuri.1@osu.edu}
\affiliation{Department of Physics and Center for Cosmology and AstroParticle Physics (CCAPP), The Ohio State University, Columbus, OH 43210, USA}

\begin{abstract}
We analyze a new class of static, smooth geometries in five-dimensional supergravity, dubbed $\mathcal{W}$-solitons. They carry the same mass and charges as four-dimensional Reissner-Nordström-like black holes but replace the horizon with a Kaluza–Klein bubble supported by electromagnetic flux. These solutions provide analytically tractable prototypes of black hole microstates in supergravity, including a new, relevant neutral configuration involving a massless axion field. Focusing on photon scattering and scalar perturbations, we compute their key observables, aiming to identify mesoscopic observables. We find that $\mathcal{W}$-solitons feature a single photon sphere, qualitatively similar to that of the black hole but with quantitative differences. They have only short-lived quasinormal modes~(QNMs), as black holes, while long-lived echo modes seen in other ultracompact horizonless objects are absent. As a result, the ringdown closely resembles that of a black hole while still showing sizable deviations. The latter are at the ${\mathcal{O}}(10\%)$ level, compatible with the recent  measurement of GW250114 and potentially falsifiable in the near future. Finally, we show that $\mathcal{W}$-solitons are stable under scalar perturbations. Our results underscore the qualitative similarities between $\mathcal{W}$-solitons and black holes, reinforcing their relevance as smooth black hole microstate prototypes.

\end{abstract}

\maketitle
\tableofcontents

\section{Introduction}

Unveiling the fundamental structure of black holes is one of the most compelling and active challenges in quantum gravity. Independent arguments from quantum information theory~\cite{Mathur:2009hf,Almheiri:2012rt} and from supersymmetric constructions in string theory~\cite{Mathur:2005zp,Bena:2022rna,Bena:2022ldq} both suggest that the vast microstructure underlying the Bekenstein–Hawking entropy must be encoded in the near-horizon region. This strongly points to new physics at the horizon, beyond the classical description provided by General Relativity, that could have observable consequences~\cite{Mayerson:2020tpn}. 

While generic black-hole microstates are intrinsically quantum, some can be sufficiently coherent to admit a classical description as ultra-compact solitonic objects. Such solutions are as compact as black holes but replace the horizon with smooth, horizonless geometries, thereby capturing the large-scale features of black-hole microstructure at the horizon scale.  In string theory, many such coherent microstates have been explicitly constructed for supersymmetric black holes~\cite{Heidmann:2019xrd,Bena:2022rna,Bena:2025pcy}, theorizing the only viable gravitational mechanism for sustaining smooth horizon-scale structure with a vast phase space: nontrivial topology induced by the deformation of extra compact dimensions and supported by electromagnetic flux \cite{Gibbons:2013tqa}.
A key ingredient in these constructions is the interplay between extra dimensions and nontrivial topological structures, which enables smooth caps that resolve the curvature singularities of black holes at their horizon. Although microstate geometries are typically formulated within the ten- or eleven-dimensional frameworks of string theory, consistent dimensional reductions often retain the essential physical features --~such as smoothness, conserved charges, and nontrivial topology. In particular, five-dimensional supergravity has proven to be a fruitful arena for constructing regular, asymptotically S$^1\times\mathbb{R}^{1,3}$ solitonic solutions~\cite{Bena:2007kg,Heidmann:2021cms,Chakraborty:2025ger}.

While supersymmetric microstate geometries have been central in probing black-hole physics beyond General Relativity \cite{Tyukov:2017uig,Bena:2020iyw,Bena:2019azk,Martinec:2020cml,Bena:2020yii,Chakrabarty:2021sff,Bianchi:2022qph}, supersymmetric models fall short of describing realistic astrophysical black holes. 
The advent of black-hole astronomy is providing unprecedented opportunities to probe the nature of compact objects. Recent LIGO–Virgo–KAGRA spectroscopy measurements~\cite{LIGOScientific:2025epi,LIGOScientific:2025obp,Berti:2025hly}, together with future prospects from LISA~\cite{LISA:2024hlh,Barausse:2020rsu,Datta:2019epe,Bhagwat:2021kwv,Maggio:2021uge}, the Einstein Telescope~\cite{Abac:2025saz,Bhagwat:2023jwv}, and the next-generation Event Horizon Telescope~\cite{Bailes:2021tot,Haworth:2019urs,Ayzenberg:2023hfw}, provide new channels to test the dynamics of spacetime in the strong-field regime. In this context, it becomes essential to construct and analyze coherent microstate geometries of astrophysically relevant, nonextremal black holes. Achieving this goal requires addressing the full nonlinear structure of Einstein’s equations within supergravity.

Recent progress has demonstrated that this program can be carried out. In particular, the development of integrable methods in supergravity~\cite{Bah:2021owp,Heidmann:2021cms,Chakraborty:2025ger} has led to the explicit construction of nonextremal topological solitons, including coherent microstates of the Schwarzschild black hole~\cite{Bah:2022yji,Bah:2023ows} and the so-called topological star~\cite{Bah:2020ogh,Bah:2020pdz}. The latter is especially notable for its analytic tractability and its reliance on the essential ingredients for horizon-scale microstructure — nontrivial topology and flux. Nevertheless, it suffers from key limitations: it lacks a spin structure,\footnote{The spin structure of a classical geometry specifies the data required to define spinor fields (e.g., Dirac fields) globally on the manifold, and is therefore a \emph{sine qua non} for any gravitational configuration compatible with fermions~\cite{Geroch:1968zm,Hawking:1977ab}.} much like the vacuum bubble-of-nothing \cite{Witten:1981gj}, and it differs significantly from nonextremal black holes due to $\mathcal{O}(1/r)$ deviations in the asymptotic metric.

In this work, we investigate a novel family of nonextremal, static, smooth horizonless solutions recently constructed in five-dimensional supergravity, known as \emph{$\mathcal{W}$-solitons} \cite{Chakraborty:2025ger}. These configurations are asymptotic to four-dimensional Minkowski spacetime times a compact circle and carry the same mass and charges as Reissner–Nordström-like black holes of four-dimensional supergravity \cite{Chow:2014cca}, including the astrophysically relevant neutral case. Their smooth caps are supported by a topological bubble threaded by electromagnetic flux. Like the topological star, they offer analytic tractability, but they overcome its main deficiencies, thereby offering excellent and analytically tractable prototypes of nonextremal black hole microstates. In particular, their neutral limit yields a new smooth horizonless geometry that can be compared to the Schwarzschild black hole, and which, from a four-dimensional perspective, is supported by a massless axion field and the usual dilaton arising from compactification.

We derive the observational and dynamical properties of $\mathcal{W}$-solitons and compare them with those of classical black holes. We focus on three physical diagnostics: (i) the structure of null geodesics and the associated photon sphere, (ii) gravitational lensing and imaging through ray-tracing simulations, and (iii) the dynamics of linear scalar perturbations, including QNM spectra~\cite{Berti:2009kk} and time-domain, ringdown responses~\cite{Berti:2025hly}. We find that $\mathcal{W}$-solitons reproduce several hallmarks of black-hole phenomenology: they support a single unstable photon sphere and short-lived QNMs without late-time echoes \cite{Cardoso:2016rao, Cardoso:2017cqb}. At the same time, quantitative sizable deviations persist, which could serve as observational smoking guns in both gravitational-wave and electromagnetic signals.
Notably, a $\mathcal{W}$-soliton arises as a fundamental entity in a fixed low-energy limit of string theory, and therefore contains no freely tunable parameters.
As a result, the solution exhibits a distinctive signature relative to a traditional black hole.\footnote{This contrasts with other phenomenological models of exotic compact objects, which typically involve extra parameters that can be tuned to reproduce black-hole properties arbitrarily well, so that observations can at best place upper bounds on the model~\cite{Cardoso:2019rvt}.}

\section{$\mathcal{W}$-solitons and black holes}
In~\cite{Chakraborty:2025ger}, a new solution-generating technique in $\mathcal{N}=2$ five-dimensional supergravity coupled to two vector multiplets enabled the construction of static nonextremal black string solutions and novel smooth horizonless topological solitons. These solutions are asymptotically four-dimensional Minkowski spacetime times a compact extra dimension,\footnote{For this reason, a black-object solution can be referred to either as a black hole or as a black string, depending on whether one takes a four-dimensional or five-dimensional perspective. In what follows, we will use both terms interchangeably.} and they can coexist within the same regime of mass and electromagnetic charges.

In this paper, we focus on the class of smooth horizonless solutions, the $\cW$-solitons, and their associated black string counterparts. For simplicity, we consider a truncation to five-dimensional minimal supergravity, which corresponds to Einstein–Maxwell gravity with a single $U(1)$ gauge field and a Chern–Simons term:
\begin{equation}
\mathcal{S} =  \frac{1}{16 \pi G_5} \int \left( R_5 \star_5 1 
\!-\! \frac32 \,F \!\wedge\! \star_5 F \!-\! F \!\wedge\! F \!\wedge\! A \right),
\label{eq:L5min}
\end{equation}
where $G_5$ is the five-dimensional Newton constant, $F = dA$ is the field strength of the gauge field $A$, and $R_5$ is the Ricci scalar. This theory arises from the compactification of M-theory on a rigid six-dimensional torus.

We decompose the spacetime into a four-dimensional part parametrized by $(t, r, \theta, \phi)$ in spherical coordinates, and a compact fifth dimension $\psi$. The periodicity lattice of the compact direction is
\begin{equation}
    (\psi,\phi)=(\psi,\phi)+(0,2\pi),\qquad (\psi,\phi)=(\psi,\phi)+(2\pi R_\psi,0),
    \label{eq:Periodicities}
\end{equation}
where $R_\psi$ denotes the asymptotic radius of the compact circle.

\subsection{The $\mathcal{W}$-soliton}
\label{sec:WSolGen}

The $\cW$-solitons form a class of static, nonextremal, smooth, horizonless geometries that possess the same mass and charges as four-dimensional black holes but resolve the horizon in five dimensions through a smooth Kaluza-Klein bubble supported by electromagnetic flux. The solution is dyonic, carrying both electric and magnetic charges under the $U(1)$ gauge field, as well as electric and magnetic Kaluza–Klein charges along the compact fifth dimension. We consider the special case where all charges have equal magnitude, as constructed in~\cite{Chakraborty:2025ger}. The solution is described by the following metric and $U(1)$ gauge field:
\begin{equation}
\begin{split}
    ds^2_\cW &= \frac{f_Q(2f_M-f_Q)}{f_M^2} \left(d\psi+\frac{Q\,f_M}{r\,f_Q} dt+Q(\cos \theta+1) \,d\phi \right)^2 - \frac{f_M}{f_Q}\,dt^2+ f_M \left[\frac{dr^2}{2f_M-f_Q}+r^2 \left( d\theta^2 +\sin^2 \theta\,d\phi^2 \right) \right], \\
    A &= \frac{Q}{r}\,dt+ \frac{f_Q-f_M}{f_M}\left(d\psi+Q (\cos \theta+1) \,d\phi \right)- Q (\cos \theta+1) \,d\phi\,,
    \label{eq:W_sol_metric}
\end{split}
\end{equation}
with
\begin{equation}
    f_M \equiv 1-\frac{2M}{r},\qquad f_Q \equiv 1-\frac{2Q^2}{r^2},
\end{equation}
where $M$ and $Q$ denote the mass and the magnitude of the charge, respectively (in units with $G_4 = 1$).\footnote{See Section 4.3 of~\cite{Chakraborty:2025ger} for more details on the charge content.}

Strictly speaking, the coordinates $(r,\theta,\phi)$ are not standard spherical coordinates, because the two-sphere has radius $r\sqrt{f_M}$ rather than $r$. One can introduce proper spherical coordinates $(R,\theta,\phi)$ with $R=r\sqrt{f_M}$; the corresponding form of the solution is given in Appendix \ref{app:SpherCoorWSol}. However, we prefer the $(r,\theta,\phi)$ system in the main text, since, in these variables, the fields are rational functions of the radial coordinate, making analytic manipulations more convenient.

The geometry terminates at the coordinate degeneracy of the $\psi$ direction, where $2f_M - f_Q = 0$. This occurs at $r = r_+$, the largest of the two roots:
\begin{equation}
    r_\pm \equiv 2 M \pm \sqrt{2(2 M^2 - Q^2)}.
    \label{eq:RpmDef}
\end{equation}
 Hence, the domain of existence of the soliton requires: 
\begin{equation}
    |Q|\leq \sqrt{2}M
    \,,
    \label{eq:ValidityBound}
\end{equation}
with equality corresponding to the extremal and supersymmetric limit~\cite{Chakraborty:2025ger}.

Regularity at the coordinate degeneracy (ensuring that the local topology is a smooth $\mathbb{Z}_k$ quotient of $\mathbb{R}^2$), together with the quantization of the Kaluza–Klein monopole charge at the poles of the two-sphere, imposes the following conditions~\cite{Chakraborty:2025ger}:\footnote{Regularity at the coordinate degeneracy requires the local $(\rho,\psi)$ subspace to take the form $ds_{\rho\psi}\propto d\rho^2 +\frac{\rho^2}{k^2 R_\psi^2} d\psi^2$, where $\rho^2 = 4(r - r_+)$, ensuring that the $(\rho,\psi)$ plane is locally a smooth $\mathbb{R}^2/\mathbb{Z}_k$ orbifold. Additionally, the quantization of the Kaluza–Klein monopole charge requires that the degeneracy of the $\phi$-circle at $\theta = 0$ is consistent with the $2\pi$ periodicity of the $\phi$ coordinate.}
\begin{equation}
    k R_\psi = \frac{r_+}{\sqrt{2}},\qquad Q= \frac{N}{2} R_\psi\,,\qquad (N,k)\in \mathbb{Z},\quad k\geq 1,
\end{equation}
where $R_\psi$ is the radius of the fifth dimension, defined by the periodicity of the coordinate $\psi = \psi +2\pi R_\psi$.

These conditions can be inverted if and only if $2k \geq |N|$, yielding a quantized family of smooth solitons with:
\begin{equation}
    M = \frac{4k^2+N^2}{8\sqrt{2} k}\,R_\psi,\qquad Q = \frac{N}{2}\,R_\psi\,,\qquad 2k \geq |N|.
    \label{eq:QuantizationRelation}
\end{equation}
Both the mass and the charge can decouple from the radius of the fifth dimension, $R_\psi$, by fine-tuning the integers $k$ and $N$, with the lower bound $M\geq \frac{R_\psi}{2\sqrt{2}}$. In particular, if $k\gg 1$, the mass $M$ can be parametrically larger than $R_\psi$.
The extremal limit, corresponding to $2k = |N|$ where $|Q|=\sqrt{2}M$, transitions the end-of-spacetime topology from a bolt ($\mathbb{R}^2/\mathbb{Z}_k$) to a NUT center ($\mathbb{R}^4/\mathbb{Z}_{|N|}$). As shown in~\cite{Chakraborty:2025ger}, the extremal solution corresponds to a spectral flow of a supersymmetric Gibbons-Hawking center which forms the basic building blocks of multicenter bubbling geometries that realize supersymmetric black hole microstates in supergravity~\cite{Bena:2007kg,Heidmann:2017cxt,Bena:2017fvm,Warner:2019jll}.\\

The size of the soliton, namely the radius of the two-sphere at the bubble locus $r = r_+$, is given by
\begin{equation}
    R_\cW = r \sqrt{f_M}|_{r=r_+}= \sqrt{r_+(r_+-2M)} = \frac{\sqrt{4k^2-N^2}}{2}\,R_\psi,
    \label{eq:SolitonSize}
\end{equation}
which reduces to $2\sqrt{2}M$ in the neutral limit and vanishes in the extremal limit, where the whole sphere smoothly shrinks. Thus, the non-extremal $\cW$-soliton exhibits the unconventional but essential property of gravitational topological solitons: its size scales naturally with the ADM mass and Newton's constant, much like black holes, and in contrast to many ultra-compact objects supported by exotic matter~\cite{Cardoso:2019rvt,Alho:2023mfc}.  

The $\cW$-soliton shares many similarities with the topological star of~\cite{Bah:2020ogh,Bah:2020pdz}: both are static, non-extremal, smooth, and horizonless solitons in a five-dimensional theory of gravity coupled to a Maxwell field. However, the $\cW$-soliton exhibits several advantages:
\begin{itemize}
\item While the bolt in the topological star has the structure of a Euclidean Schwarzschild geometry, the bolt of the $\cW$-soliton carries Kaluza–Klein monopole charge and is locally a Taub-bolt geometry, which admits a spin structure \cite{Bossard:2014yta}.
\item The extremal limit of the topological star is singular (corresponding to a singular extremal black string), whereas the extremal limit of the $\cW$-soliton remains smooth and horizonless, reducing to a supersymmetric Gibbons–Hawking center.
\item The metric component along the fifth dimension in the topological star decays as $1+\mathcal{O}(r^{-1})$ at large distances, leading to a mismatch between the five-dimensional and four-dimensional mass. In contrast, the $\cW$-soliton features no $1/r$ term in the large-$r$ expansion of $g_{\psi\psi} \sim 1 - \frac{4M^2}{r^2} + \dots$, ensuring that the five- and four-dimensional masses coincide.
\end{itemize}

Finally, the neutral limit $N = Q = 0$ of the $\cW$-soliton is particularly interesting. It does not reduce to the standard Euclidean Schwarzschild geometry plus a time, also called vacuum Kaluza–Klein bubble or bubble-of-nothing \cite{Witten:1981gj}, but instead yields a new smooth horizonless solution and a nontrivial axion profile:
\begin{equation}\label{eq:neutral_W}
\begin{split}
     ds^2_{\mathcal{W}}|_{Q =0} &= -\left( 1 - \frac{2M}{r} \right) dt^2 + \frac{r-2M}{r-4M} dr^2 +\frac{r(r-4M)}{(r-2M)^2} d\psi^2 + r(r-2M)(d\theta^2 +\sin^2 \theta \, d\phi^2),\\
    A &= \frac{2M}{r-2M} d\psi \,,
\end{split}
\end{equation}
and the spacetime smoothly ends\footnote{The point $r=4M$ is regular and plays the same role as the standard origin ($r=0$) describing the center of symmetry of a spherically symmetric star. As we shall discuss, geodesics and perturbations are ``reflected'' there in the same sense they are reflected at the center of symmetry of a star in spherical coordinates.} at $r=4M$. The solution agrees with the Schwarzschild black-hole metric at large distance (up to ${\cal O}(M^2/r^2)$ terms) and
has only one free parameter (the mass $M$), just like the Schwarzschild solution.

\subsection{The black hole}

The $\cW$-soliton exists in the same regime of mass and electromagnetic charges as a non-extremal black string in five-dimensional minimal supergravity, which, upon Kaluza-Klein reduction along the fifth compact dimension, yields a static charged black hole. One might initially expect this black string to arise from the five-dimensional embedding of the four-dimensional dyonic Reissner–Nordström black hole.\footnote{The embedding of the dyonic Reissner–Nordström black hole in five-dimensional minimal supergravity is given by $A = - \frac{Q}{r} dt - P (\cos \theta+1) \,d\phi$ and
\begin{equation}
ds_\text{RN}^2 = \left( d\psi +Q (\cos \theta+1) \,d\phi-\frac{P}{r} dt\right)^2 + \left(1-\frac{2M}{r} +\frac{P^2+Q^2}{r^2} \right) dt^2 + \frac{dr^2}{1-\frac{2M}{r} +\frac{P^2+Q^2}{r^2}}+r^2 d\Omega_2^2\,,\nonumber
\end{equation}} However, due to differences in the relative signs of the electric and magnetic charges, the $\cW$-soliton does not share the same conserved charges as this black hole.

The appropriate black string to compare with has been derived in~\cite{Chakraborty:2025ger}, and is given by:
\begin{equation}
\begin{split}
    ds^2_{\rm BH} &= f_+ f_- \left( d\psi + \frac{Q}{r\,f_+} dt + Q (\cos \theta+1) \, d\phi \right)^2-\frac{f_M}{f_+ f_-}\,dt^2 + f_-\left[ \frac{dr^2}{f_- f_M} + r^2 (d\theta^2 + \sin^2\theta \, d\phi^2)\right] \,, \\
A &= \frac{Q}{r}dt + \frac{f_+-f_-}{2}\left(d\psi + Q (\cos \theta+1)\, d\phi\right) - Q (\cos \theta+1) \,d\phi \,. \label{eq:BlackString}
\end{split}
\end{equation}
where we have defined:
\begin{equation}
    f_\pm \equiv 1 \pm \frac{\sqrt{M^2+4Q^2}-M}{r}.
\end{equation}
As for the $\cW$-soliton, the radial variable $r$ does not correspond to the genuine spherical radial coordinate as it does not coincide with the radius of the S$^2$. The actual radial coordinates $(R,\theta,\phi)$ require the change
\begin{equation}
    r\sqrt{f_-} = R \qquad \Leftrightarrow \qquad r=\sqrt{\left(M - \sqrt{M^{2} + 4Q^{2}}\right)^{2} + 4R^{2}} +\sqrt{M^{2} + 4Q^{2}}- M. \label{eq:TrueSpherBS}
\end{equation}

The black string has the same conserved charges as the $\cW$-soliton with the correct signs. Moreover, the regularity at the poles of the S$^2$ imposes the same quantization of the Kaluza–Klein monopole charge $Q=NR_\psi/2$ as for the $\cW$-soliton~\eqref{eq:QuantizationRelation}. However, the mass is completely decoupled from the extra dimension size.

It has a horizon at $r=2M$ for any $Q$, with a S$^1\times$S$^2$ topology, typical of a black string. The entropy and temperature are given by (in units $G_4=1$)
\begin{equation}
\begin{split}
S= \pi \left(M+\sqrt{M^2+4Q^2} \right)^\frac{1}{2}\left(3M-\sqrt{M^2+4Q^2} \right)^\frac{3}{2},\qquad T = \frac{1}{4\pi \sqrt{\left(M+\sqrt{M^2+4Q^2}\right)^2 -8Q^2}}.
\end{split}
\end{equation}
Requiring that the horizon appears before the singularities at the zero of $f_-$, imposes an extremal bound identical to the $\cW$-soliton, Eq.~\eqref{eq:ValidityBound}. Thus, for a given black string at a given $(M,Q)$ satisfying~\eqref{eq:ValidityBound}, there exists a $\cW$-soliton that has replaced the horizon by a smooth horizonless topological structure supported by electromagnetic flux. 

At extremality, $M=|Q|/\sqrt{2}$, both the black string and the $\cW$-soliton become the same smooth horizonless supersymmetric center.

Moreover, one can show that the specific heat $\frac{\partial M}{\partial T}|_Q$ is always negative for $|Q|<\sqrt{2}M$ indicating that the black string is thermodynamically unstable as the Reissner-Nordström black hole.

Finally, the neutral limit $Q = 0$ yields the trivial embedding of the Schwarzschild black hole in five dimensions:
\begin{align}\label{eq:neutral_BS}
    ds^2_{\rm BH}|_{Q = 0} &= -\left( 1-\frac{2M}{r} \right) dt^2 + \left( 1-\frac{2M}{r} \right)^{-1} dr^2 + d\psi^2 + r^2 (d\theta^2 + \sin^2 \theta \, d\phi^2) \,,\\ 
    A&=\, 0.
\end{align}

\subsection{Four-dimensional profiles}

Despite being constructed in five dimensions, the independence from the compact fifth dimension, which decouples at the boundary \eqref{eq:Periodicities}, allows for a four-dimensional description of the solution via dimensional reduction along the S$^1$. The dimensional reduction of $\mathcal{N}=2$ five-dimensional supergravity coupled to two vector multiplets yields the so-called four-dimensional ``STU'' supergravity, consisting of three complex scalars, $z^I$, and four U(1) gauge fields, $(\mathcal{A},A^{I})$ (see, for example, \cite{DallAgata:2010srl,Chow:2014cca}). In minimal supergravity, the index $I$ drops, so the three scalars and three gauge fields become respectively identical.

From the four-dimensional viewpoint, the $\cW$-soliton is described by\footnote{The compactification steps are discussed in detail in section 2.6.3 of \cite{Chakraborty:2025ger}}:
\begin{equation}
\begin{split}
    ds^2_\cW &= \sqrt{f_Q(2f_M-f_Q)}\left[ - \frac{dt^2}{f_Q}+ \frac{dr^2}{2f_M-f_Q}+r^2 \left( d\theta^2 +\sin^2 \theta\,d\phi^2 \right) \right], \\
    A &= \frac{Q}{r}\frac{f_M}{f_Q}\,dt- Q (\cos \theta+1) \,d\phi\,, \quad \mathcal{A} = -\frac{Q\,f_M}{r\,f_Q} dt-Q(\cos \theta+1) \,d\phi, \quad z = \frac{f_Q-f_M}{f_M} \,+\, i \, \frac{\sqrt{f_Q(2f_M-f_Q)}}{f_M}\,,
    \label{eq:W_sol_metric4d}
\end{split}
\end{equation}
and the neutral limit can be easily obtained by setting $Q=0$. In this limit, the two gauge fields vanish identically, and one is left with a nontrivial complex scalar, whose real and imaginary parts correspond to the axion and dilaton fields, respectively.
As expected, the four-dimensional solution features a singularity at the bubble locus where $2f_M-f_Q=0$. This behavior is characteristic of geometries reduced along a smoothly degenerating compact direction: while the higher-dimensional geometry is regular, the coordinate degeneration manifests as a naked singularity from the lower-dimensional perspective.

The black string reduces to a regular dyonic and dilatonic black hole in four dimensions:
\begin{equation}
\begin{split}
    ds^2_{\rm BH} &= -\frac{f_M}{\sqrt{f_+ f_-}}\,dt^2 + \sqrt{f_+ f_-}\left[ \frac{dr^2}{f_M} + r^2 f_- (d\theta^2 + \sin^2\theta \, d\phi^2)\right] \,, \\
A &= \frac{Q}{r \,f_+}dt - Q (\cos \theta+1) \,d\phi \,, \qquad \mathcal{A}=-\frac{Q}{r\,f_+} dt - Q (\cos \theta+1) \, d\phi, \qquad z= \frac{f_+-f_-}{2}\,+\, i \,\sqrt{f_+ f_-} .\label{eq:BlackString4d}
\end{split}
\end{equation}
This black hole solution closely resembles the dyonic Reissner–Nordström black hole. However, a sign difference in the electric charges of the Kaluza–Klein vector, $\mathcal{A}$, induces a subtle difference that requires the presence of a nontrivial complex scalar. As a result, the vector fields cannot be combined into a single one, as for the Reissner–Nordström solution.

\section{Photon scattering and photon sphere}

We now analyze the physics of null geodesics in the backgrounds introduced above. Geodesics are parametrized by the affine parameter $\tau$, and we define the conjugate momenta $p_\mu$ as
\begin{equation}
    p_\mu \equiv g_{\mu \nu} \,\dot{x}^\nu\,,
\end{equation}
where $\dot{x}^\nu = \frac{dx^\nu}{d\tau}$.
For spacetimes with three isometries along $(t, \phi, \psi)$, null geodesics admit four conserved quantities: the Hamiltonian $H = \frac{1}{2} g^{\mu\nu} p_\mu p_\nu = 0$, and the momenta associated with the Killing vectors, given by $p_t = -E$, $p_\phi$, and $p_\psi$.
The radial and angular dynamics of the geodesics, described by the functions $r(\tau)$ and $\theta(\tau)$, can be derived from the Hamilton equation:
\begin{equation}
    \dot{p}_\mu = -\frac{dH}{dx^\mu}\,.
\end{equation}
Moreover, the symmetric structure of the backgrounds guarantees the existence of a Carter constant that enables separation of variables. For both the $\mathcal{W}$-soliton and the black string, we find that this constant takes the form
\begin{equation}
    C = p_\theta^2+ \frac{1}{\sin^2 \theta} \left[\left(p_\phi-2p_\psi Q\right)^2 \cos^2 \frac{\theta}{2}+ p_\phi^2 \sin^2 \frac{\theta}{2} \right].
\end{equation}
In what follows, we focus on closed null orbits and their properties. Specifically, we restrict to effectively massless four-dimensional probes, corresponding to the lowest Kaluza–Klein modes with $p_\psi = 0$.\footnote{Note that probes with no momentum along $\psi$ can still exhibit a nonzero $d\psi/dt$ due to the presence of a nonvanishing $g_{t\psi}$ component in the background metric. This effect is analogous to frame dragging in rotating geometries, except that here the ``rotation'' occurs along the fifth dimension.} Indeed, $p_\psi$ generates an effective mass term in the Hamiltonian (and also effective charges under the Kaluza-Klein vector field): $g^{\psi\psi} p_\psi^2 \to p_\psi^2$ asymptotically.

\subsection{Geodesic equations}

First, we find that null geodesics in the $\mathcal{W}$-soliton background~\eqref{eq:W_sol_metric} obey the following first-order equations:
\begin{align}
    \label{eq:NullGeo}
    \dot{t}&= \frac{f_Q}{f_M}\,E+\frac{Q\,p_\psi}{r},\qquad \dot{\phi}= \frac{p_\phi-Q\,p_\psi(1+\cos \theta)}{r^2 f_M \sin^2 \theta},\qquad \dot{\psi}= \frac{f_M^2}{f_Q(2f_M-f_Q)}\,p_\psi-\frac{Q f_M}{r f_Q} \,\dot{t}-Q(1+\cos \theta) \,\dot{\phi}, \nonumber \\
    \dot{\theta} &= \pm \frac{\sqrt{\Theta(\theta)}}{r^2 f_M},\qquad \dot{r}= \pm\sqrt{R(r)},
\end{align}
where we have defined
\begin{equation}
\begin{split}
    \Theta(\theta) &\equiv C -\frac{1}{\sin^2 \theta} \left[\left(p_\phi-2p_\psi Q\right)^2 \cos^2 \frac{\theta}{2}+ p_\phi^2 \sin^2 \frac{\theta}{2} \right],\\
    R(r) &\equiv \frac{2f_M-f_Q}{r^2 f_M^2 f_Q} \left[(r E f_Q + Q p_\psi f_M)^2+\left(Q^2 f_Q -\frac{r^2 f_M(f_M-f_Q)^2}{2 f_M-f_Q}\right)\,p_\psi^2-C f_Q \right]
    \label{eq:Theta-R}
\end{split}
\end{equation}

Second, we find that the null geodesics for the black-string geometry~\eqref{eq:BlackString} obey:
\begin{align}
    \dot{t}&= \frac{f_-}{f_M}\left( f_+\,E+\frac{Q}{r}\,p_\psi\right),\qquad \dot{\phi}= \frac{p_\phi-Q\,p_\psi(1+\cos \theta)}{r^2 f_- \sin^2 \theta},\qquad \dot{\psi}= \frac{1}{f_- f_+}\,p_\psi-\frac{Q}{r f_+} \,\dot{t}-Q(1+\cos \theta) \,\dot{\phi}, \nonumber \\
    \dot{\theta} &= \pm \frac{\sqrt{\Theta(\theta)}}{r^2 f_-},\qquad \dot{r}= \pm\sqrt{\widetilde{R}(r)},
    \label{eq:NullGeoBS}
    \end{align}
where $\Theta$ is the same as in~\eqref{eq:Theta-R} and $\widetilde{R}$ is given by
\begin{equation}
    \widetilde{R}(r) \equiv \frac{f_M}{r^2 f_-} \left[\frac{r^2 f_+ f_-^2}{f_M} \left( E+\frac{Q}{r f_+} p_\psi \right)^2- \frac{r^2-Q^2 f_+}{f_+} p_\psi^2-C \right].
    \label{eq:RdefBS}
\end{equation}
We are free to choose $E=1$, which corresponds to taking the affine parameter $\tau$ as the proper time of the probe.

\subsection{Photon sphere and Lyapunov exponents}

\subsubsection{For the $\cW$-soliton}

Photon trajectories in the $\cW$‑soliton background obey the null‑geodesic equations~\eqref{eq:NullGeo} with the additional requirement that the probe is effectively massless in four dimensions, i.e. $p_\psi = 0$. Probes that do not have dynamics along the fifth dimension experience only a spherically-symmetric spacetime and the angular equation reduces to its usual spherically symmetric form. Thus, without loss of generality, we may restrict attention to motion in the equatorial plane, $\theta = \pi/2$.\footnote{In the equatorial plane, the Carter constant is fixed in terms of the probe’s angular momentum as $C = p_\phi^{2}$.}

A photon sphere is a closed null orbit of constant radial coordinate $r = R_o$, obtained by demanding $\dot r = \ddot r = 0$. The first condition yields two roots outside the bubble locus $r_+$ (one of which coincides with it), but imposing $\ddot r = 0$ fixes $p_\phi$ and forces those two roots to coincide. Hence, the $\cW$‑soliton possesses a single photon orbit, located precisely at the bubble locus $R_o = r_+$~\eqref{eq:RpmDef}. Its characteristics are
\begin{equation}
    R_o=r_+,\qquad p_\phi=\Omega_\cW^{-1} =\sqrt{2}\,\lambda_\cW^{-1}= \sqrt{2r_+ (r_+-2M)}, 
    \label{eq:PhotonOrbitPropWSol}
\end{equation}
where $\Omega_\cW=\dot{\phi}/\dot{t}$ is the angular velocity at the orbit, and $\lambda_\cW$ is the Lyapunov exponent measuring the average rate of expansion or contraction of adjacent geodesics at the photon sphere given by, for spherically symmetric spacetimes~\cite{Cornish:2003ig,Cardoso:2008bp}, 
\begin{equation}
    \lambda = \sqrt{\frac{1}{2 \dot{t}^2}\,\frac{d^2 R(r)}{dr^2}} \,\,\Bigl|_{r=R_o}\,,
    \label{eq:LyapExp}
\end{equation}
The fact that $\lambda$ is real implies the orbit is \emph{unstable}. 

Note that $R_o$ does not represent the physical size of the photon sphere, as the radial coordinate $r$ does not coincide with the proper radius of the two-sphere. The physical radius of the two-sphere at the photon orbit is \eqref{eq:SolitonSize}:
\begin{equation}
    R_{\cW} = \sqrt{r_+ (r_+-2M)}= \frac{1}{\lambda_\cW}.
\end{equation}

In the neutral limit ($Q = 0$) one finds
\begin{equation}
    R_O = 4M,\qquad p_\phi=\Omega_\cW^{-1}=4M,\qquad \lambda_\cW = {R_{\cW}}^{-1}= \frac{1}{2\sqrt{2}\,M},
\end{equation}
which are similar to the corresponding values of the Schwarzschild photon sphere; the detailed comparison is postponed to Sec.~\ref{sec:ComparisonPhotonSpheres}.

To conclude, the $\cW$‑soliton shares similarities in photon scattering properties with the topological star, whose photon spheres were analyzed in~\cite{Heidmann:2022ehn}. For certain mass–charge regimes, a topological star features a single unstable photon orbit at the bubble locus (a topological star of the first kind), while in other regimes, it admits a second, stable inner photon sphere (a topological star of the second kind). The $\cW$‑soliton always behaves as a first-kind topological star. As shown in~\cite{Heidmann:2023ojf}, the absence of a stable photon sphere prevents the formation of long-lived QNMs or echoes~\cite{Cardoso:2016rao,Cardoso:2017cqb,Dima:2024cok,Bena:2024hoh,Dima:2025zot}. We therefore expect the $\cW$‑soliton to exhibit only short-lived QNMs, with no late-time echoes, much like an ordinary black hole. We will discuss this point in detail in Sec.~\ref{sec:timedomain}.

\subsubsection{For the black string}

We now apply the same analysis to the black string geometry, for which the radial potential of null geodesics is given by~\eqref{eq:RdefBS}. Again, we restrict to effectively massless probes ($p_\psi = 0$) and consider motion in the equatorial plane ($\theta = \pi/2$), which is justified by spherical symmetry. We search for closed null orbits at constant radial coordinate $r = R_o$, imposed via $\dot r = \ddot r = 0$.

We find that the black string admits a unique photon sphere outside its horizon. After some algebra, we get:
\begin{align}
    R_o &= \frac{\left(\sqrt{m_+}+\sqrt{9m_+-16m_-} \right)\left(3\sqrt{m_+}+\sqrt{9m_+-16m_-} \right)}{8}, \nonumber \\
    p_\phi &= \Omega_\text{BH}^{-1}=\frac{\left(\sqrt{9m_+-16m_-}-\sqrt{m_+} \right)^\frac{1}{2} \left(3\sqrt{m_+}+\sqrt{9m_+-16m_-} \right)^\frac{3}{2}}{4},\label{eq:PhotonOrbitPropBS} \\
    \lambda_\text{BH} &= \frac{4\sqrt{2}\, (9m_+-16m_-)^\frac{1}{4}}{\left(\sqrt{9m_+-16m_-}-\sqrt{m_+} \right)^\frac{1}{2} \left(3\sqrt{m_+}+\sqrt{9m_+-16m_-} \right)^2},
    \nonumber
\end{align}
where $$ m_\pm \equiv \frac{\sqrt{M^2+4Q^2}\pm M}{2}.$$

In the neutral limit $Q = 0$, so that $m_+ = M$ and $m_- = 0$, we recover the standard Schwarzschild values:
\begin{equation}
    R_o= 3M, \qquad p_\phi =\Omega_\text{BH}^{-1}= 3\sqrt{3} M, \qquad \lambda_\text{BH} = \frac{1}{3\sqrt{3} M}.
\end{equation}
 
The Lyapunov exponent $\lambda_\text{BH}$ is real for all values of $M$ and $Q$ within the validity range~\eqref{eq:ValidityBound}, confirming that the photon orbit is always \emph{unstable}.

Finally, when $Q\neq0$ the coordinate $R_o$ does not represent the physical radius of the photon sphere. The latter is given by \eqref{eq:TrueSpherBS} at $r=R_0$, that is:
\begin{equation}
    R_{\text{BH}} = r\sqrt{f_-} |_{r = R_o} = \sqrt{R_o(R_o-m_-)}.
\end{equation}

\subsubsection{Comparison}
\label{sec:ComparisonPhotonSpheres}

\begin{figure}[t]
  \centering
\includegraphics[width=0.8\textwidth]{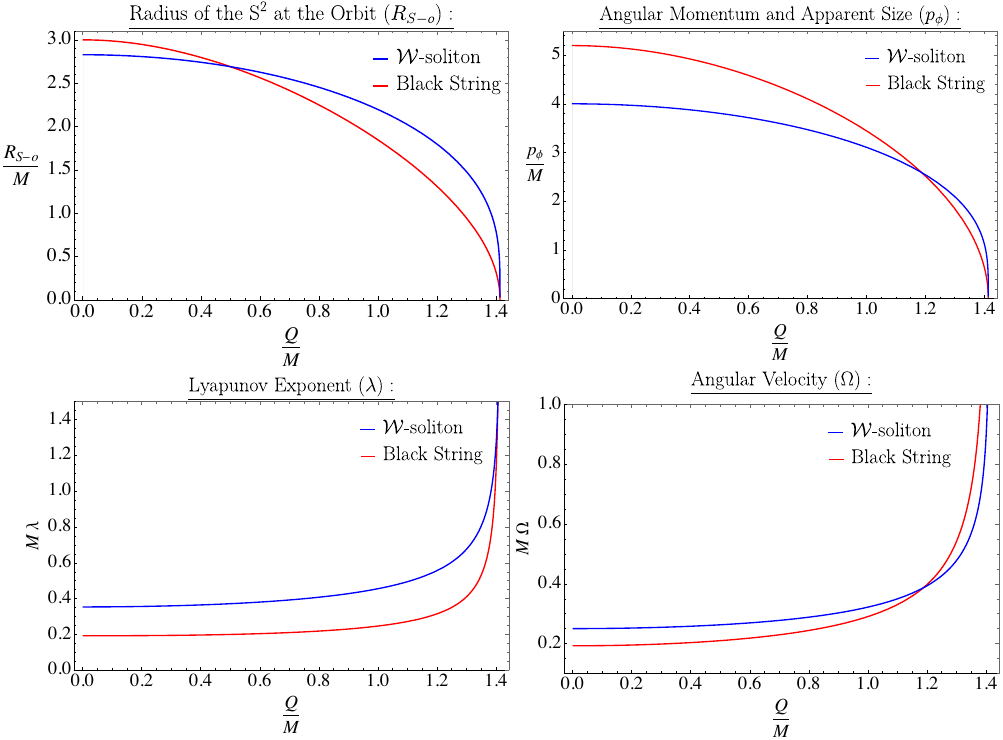}
  \caption{Properties of the photon sphere for the $\cW$-soliton and the black string as functions of the charge-to-mass ratio.
Top left panel: Radius of the round two-sphere at the location of the photon orbit. 
Top right panel: Angular momentum of photons on the orbit, which also determines the apparent size of the photon sphere as seen by an asymptotic observer.
Bottom left panel: Lyapunov exponent characterizing the instability of the photon sphere.
Bottom right panel: Angular velocity of photons at the photon sphere, related to the QNM frequencies.}
 \label{fig:PhotonSphereProp}
\end{figure}

The photon scattering properties of the $\cW$-soliton and its corresponding black string are qualitatively similar in terms of photon orbits: both spacetimes feature a single, unstable photon sphere. However, their photon spheres differ quantitatively. In particular, we observe order-one differences in how the radius of the $S^2$ at the orbit, Lyapunov exponent, angular momentum, and angular velocity depend on the charge-to-mass ratio. These dependencies are illustrated in Fig.~\ref{fig:PhotonSphereProp} and discussed in detail below.

\begin{itemize}
    \item \underline{Radius of the $S^2$ (top left panel of Fig.~\ref{fig:PhotonSphereProp})}:
The photon sphere radii for both the $\cW$-soliton and the black string are of the same order of magnitude, scale with the ADM mass, and decrease monotonically with increasing charge-to-mass ratio. At $Q = 0$, the black string's photon sphere is about 6\% larger than that of the $\cW$-soliton. However, for $Q \geq \frac{1}{2}M$, the black string's photon sphere becomes smaller. In both geometries, the radius vanishes at extremality, $Q = \sqrt{2}M$, because the solutions converge to the same smooth supersymmetric geometry ending in a NUT center, where the S$^2$ smoothly shrinks to zero.
    \item \underline{Angular momentum and apparent size (top right panel of Fig.~\ref{fig:PhotonSphereProp})}:  
The angular momentum $p_\phi$ of photons trapped on the photon sphere, which corresponds to the critical impact parameter for null geodesics arriving from infinity, follows the same qualitative behavior as the radius. It also determines the apparent size $\delta$ of the photon sphere for a distant observer via the relation:
\begin{equation}
\delta = \underset{r\to \infty}{\lim} \frac{r^2 \dot{\phi}}{\dot{r}} = p_\phi.
\label{eq:ApparentSizeGen}
\end{equation}
Consequently, the apparent size of the neutral $\cW$-soliton's photon sphere is about 15\% smaller than that of the Schwarzschild black hole. The curves cross at $Q = 1.2M$, after which the black string appears smaller. At extremality, both apparent sizes tend to zero as the geometries shrink to a point source.

\item \underline{Lyapunov exponent (bottom left panel of Fig.~\ref{fig:PhotonSphereProp})}:  
The Lyapunov exponent characterizes the instability of null geodesics at the photon sphere. Across the full range of charge-to-mass ratios, the exponent for the $\cW$-soliton is roughly twice that of the black string, indicating a significantly stronger instability. This quantity is closely linked to the decay rate of perturbations during the ringdown phase \cite{Ferrari:1984zz,Cardoso:2008bp,Heidmann:2023ojf}; therefore, we expect the ringdown signal of the $\cW$-soliton to damp approximately twice as fast as that of the black string.  

Notably, the Lyapunov exponents diverge as extremality is approached, which reflects the fact that the photon sphere collapses to a point. Intuitively, the smaller the photon sphere, the more sensitive adjacent geodesics become to initial perturbations, leading to rapid divergence in the instability.  

    \item \underline{Angular velocity (bottom right panel of Fig.~\ref{fig:PhotonSphereProp})}:  
The angular velocity of photons on the photon sphere is inversely related to their angular momentum: $\Omega = p_\phi^{-1}$. While its behavior closely tracks $p_\phi$, we plot it separately because it governs the oscillation frequency of the ringdown response to perturbations \cite{Ferrari:1984zz,Cardoso:2008bp,Heidmann:2023ojf}. As we will discuss in a later section, the real part of the QNM frequencies is approximately proportional to this angular velocity. Hence, we anticipate a $\approx15\%$ difference in oscillation frequencies between the two backgrounds in the neutral limit, with vanishing deviations occurring near $Q = 1.2M$.
\end{itemize}

Other differences in light-scattering behavior, beyond the properties of the photon sphere, also distinguish the $\cW$-soliton from the black string. For black holes, the photon sphere delineates the \emph{shadow}: a region beyond which infalling light rays are inevitably absorbed by the event horizon and cannot reach distant observers. As such, the black-string photon sphere appears entirely dark to an asymptotic observer.

In contrast, the photon sphere of the $\cW$-soliton coincides with the bubble locus. Since the soliton is horizonless, light rays reaching this surface are not absorbed at first order. Assuming no interaction with the soliton, such rays are reflected in a nontrivial manner and eventually escape to infinity.\footnote{As shown in~\cite{Heidmann:2025yzd}, this assumption may not always hold, as the soliton can induce a nontrivial tidal instability on infalling probes, potentially leading to their nonlinear disruption and trapping at the bubble locus. However, in this paper, we restrict our analysis to linear dynamics.} Consequently, instead of forming a shadow, the soliton photon sphere acts as a reflective boundary. As argued in~\cite{Heidmann:2022ehn}, this behavior causes the soliton's ``shadow'' to function as a nontrivial \emph{spacetime mirror}, in sharp contrast to the absorbing shadow of a black hole. We will investigate this further in the next section using ray-tracing simulations.

\subsection{Imaging}

In this section, we investigate the general photon scattering properties of the $\cW$-soliton and compare them to those of its corresponding black string. To this end, we numerically integrate null geodesics as seen by a distant observer using the ray-tracing techniques developed in~\cite{Heidmann:2022ehn}. We present two sets of simulations: one in an ``unrealistic'' setup designed to clearly highlight the lensing features of the geometries, and another in a more ``realistic'' astrophysical context simulating the image produced by a bright artificial accretion disk near the ISCO.

\begin{figure}[t]
  \centering
\includegraphics[width=0.54\textwidth]{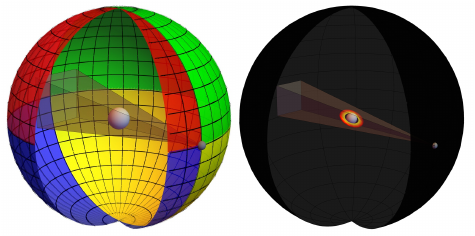}
  \caption{Illustration of the artificial background grids used for imaging. The observer (smaller gray point) is placed on a ``celestial'' sphere centered around the gravitational object (represented as a larger white sphere). For the first background, the celestial sphere is covered with a quadri-color grid of meridians and latitudes. For the second background, the sphere is entirely black except for a bright accretion disk. The disk is inclined by an angle $\pi/3$ relative to the camera-object plane and spans the radial range $[6M,7M]$. The celestial spheres have been artificially truncated here to improve visibility near the ISCO.}
  \label{fig:Methodo}
\end{figure}

The observer is placed on a ``celestial'' sphere located at a large radius $R$ (see Fig.~\ref{fig:Methodo}).\footnote{By radius we mean that the S$^2$ radius is $R$, which does not correspond to $r=R$ for our geometries since $\sqrt{g_{\theta\theta}}\neq r$.} Photons received by the observer are traced backward by integrating the geodesic equations from the observer. We construct two artificial images as seen from the observer's perspective. In the first, the celestial sphere is artificially covered with a quadri-color grid of meridians and latitudes spaced by an angle of $\pi/20$ (see the left panel of Fig.~\ref{fig:Methodo}). In the second, the celestial sphere is entirely black, and a bright source (roughly simulating an ``accretion disk'') is added with an inclination angle of $\pi/3$ relative to the camera-object plane. The disk extends radially from $6M$ to $8M$ (see the right panel of Fig.~\ref{fig:Methodo}). For the first simulation, we take $R = 15M$, and for the second, $R = 30M$. The observer is modeled as a camera with $10^6$ pixels pointing toward the center of the spacetime. Its field of view spans an angular width of $\delta \varphi = 2\pi/7$ for $R = 15M$ and $\delta \varphi = \pi/7$ for $R = 30M$. For more details about the simulation method, we refer the interested reader to \cite{Heidmann:2022ehn}.

\begin{figure}[t]
  \centering
\includegraphics[width=1\textwidth]{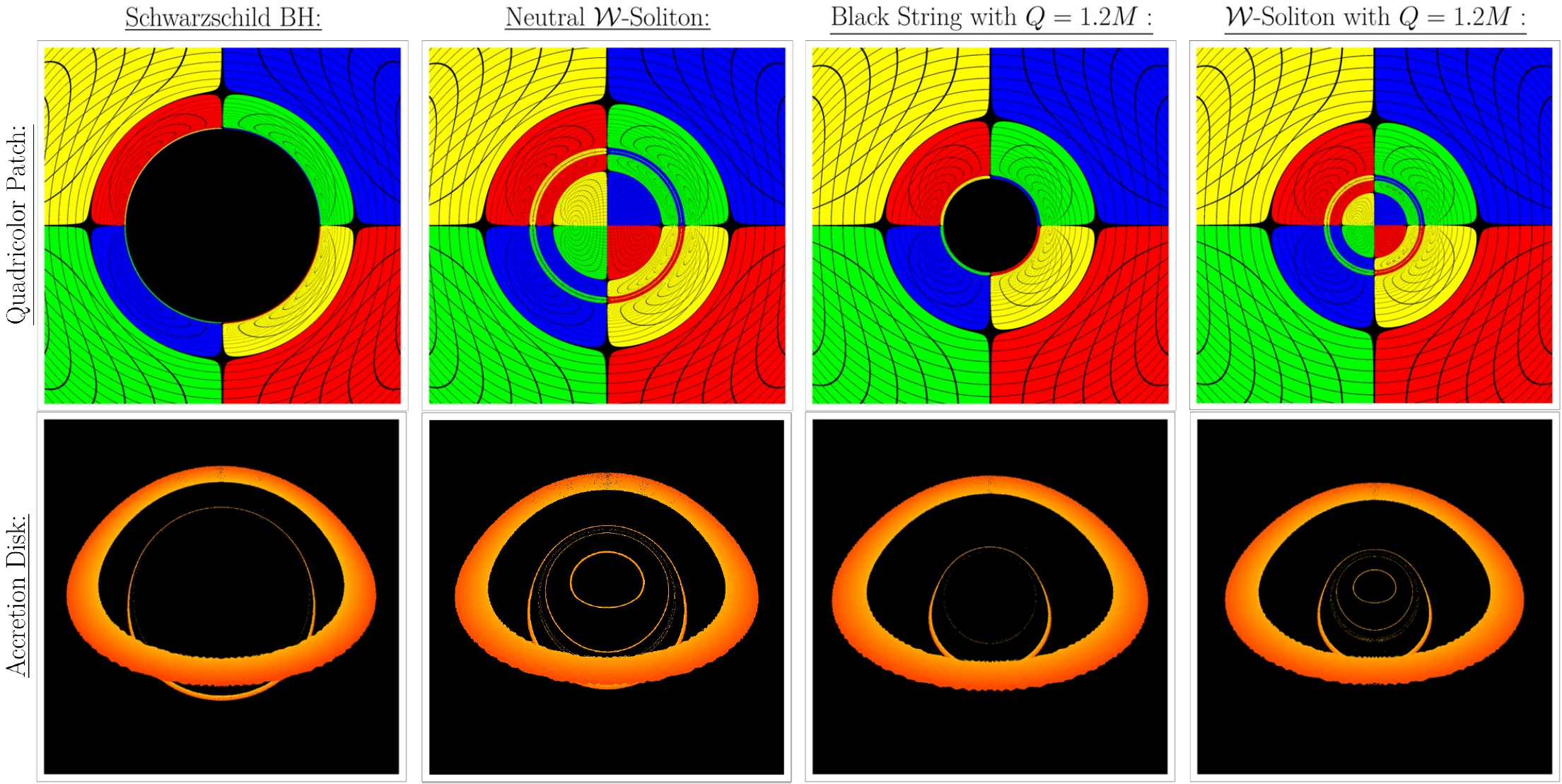}
  \caption{Gravitational lensing effects of the $\cW$-soliton compared to its corresponding black string. From left to right: the four different backgrounds for $Q=0$ and $Q=1.2M$. First row: imaging simulation with a quadri-color screen on the celestial sphere. Second row: imaging simulation of an artificial bright accretion disk near the ISCO.}
  \label{fig:QuadriColSimu}
\end{figure}

The results of the quadri-color lensing simulation are shown in the first row of Fig.~\ref{fig:QuadriColSimu}, comparing the $\cW$-soliton and the black string for two representative charge values, $Q = 0$ and $Q = 1.2M$. As expected, the photon sphere of the black string produces a shadow, i.e. a region from which no infalling light escapes to infinity, surrounded by a series of increasingly narrow ``subrings'' generated by photons orbiting the object multiple times before escaping~\cite{Johnson:2019ljv}.

The $\cW$-soliton exhibits qualitatively similar lensing features, with a single photon sphere and multiple subrings. However, because the soliton is horizonless, its photon sphere does not enclose a shadow. Instead, light rays that reach the photon sphere reflect nontrivially at the bubble locus and eventually escape. The reflection is symmetric and regular: a photon approaching the bubble surface from one direction is mirrored back along the same plane. As a result, the soliton acts as a ``spherical spacetime mirror'' for light. Despite this reflective behavior, photon trajectories experience strong bending, which is responsible for the central mirroring of the entire celestial grid seen in the soliton images in Fig.~\ref{fig:QuadriColSimu}.

As shown analytically in the previous section, the apparent size of the neutral $\cW$-soliton’s photon sphere is smaller than that of the Schwarzschild black hole. The simulation demonstrates that this reduction in size also affects the apparent size of the subrings, although the first subring remains closer in size to that of Schwarzschild. Furthermore, for $Q = 1.2M$, the simulation shows that the apparent photon sphere sizes of the $\cW$-soliton and the black string become nearly identical, with this agreement extending to the structure of the subrings. In this regime, the two geometries become nearly indistinguishable from the perspective of gravitational lensing, except for the crucial difference that the soliton replaces the black string’s dark shadow with a smooth reflective core. 

If we had increased the charge closer to extremality $Q=\sqrt{2}M$, we would have observed a smaller and smaller shadow and photon ring for the black string and soliton, respectively, as both solutions converge to the same smooth horizonless supersymmetric center. 

We now turn to the second simulation, which models a more physical scenario by introducing a mock static accretion disk near the ISCO \cite{Heidmann:2022ehn}. The corresponding results are shown in the second row of Fig.~\ref{fig:QuadriColSimu}. Despite the quantitative differences in lensing properties between the soliton and the black string, and the presence or absence of a horizon, the more realistic images appear significantly harder to distinguish.

First, neither the black string nor the soliton emits light; hence, the brightest part of the image is the direct emission from the accretion disk. This includes the portion of the disk located behind the geometry, which remains visible due to gravitational lensing around the first subring, producing the characteristic ``D-shaped'' profile. Because the first subring is comparable in apparent size for the neutral geometries, and nearly identical for the $Q=1.2M$ geometries, the direct images of the disk appear strikingly similar.

Second, there are thin indirect images of the accretion disk near the photon sphere corresponding to light rays making several turns around the photon orbit. In this region, the neutral geometries differ qualitatively, introducing subtle differences in these indirect images of the disk that require high angular resolution to detect. These features remain beyond the capabilities of the current Event Horizon Telescope, but should become accessible to next-generation space-based radio interferometers~\cite{Gralla:2020srx,Ayzenberg:2023hfw}.

Finally, the reflective nature of the soliton gives rise to additional, highly localized photon rings inside the photon sphere. These are produced by the soliton’s atypicality to preserve spherical symmetry even at the topological structure replacing the horizon. This causes light to reflect as off a mirror, generating narrow and sharply defined rings that only high-resolution instruments could detect. However, as discussed in~\cite{Heidmann:2022ehn}, for more typical topological solitons, those that break spherical symmetry close to the bubble locus, one may instead expect chaotic reflection at the bubble locus. This would result in a diffuse, low-intensity glow emanating from within the ``shadow'' region.

\section{Test-scalar perturbations and QNMs}

We now analyze the dynamics of a minimally coupled massless scalar field in the $\cW$-soliton and its corresponding black string. This analysis serves three main purposes: first, to assess the classical stability of the solutions under test scalar perturbations; second, to derive and compare the spectra of scalar QNMs for both geometries; and third, to examine and contrast their scalar ringdown responses. 

\subsection{Scalar wave equation}
\label{sec:ScalWaveGen}

A probe scalar field that is massless and minimally coupled to the background is governed by the Klein–Gordon equation:
\begin{equation}
    \frac{1}{\sqrt{-g}} \partial_\mu(g^{\mu\nu} \sqrt{-g} \, \partial_\nu \Phi)=0 \,,
    \label{eq:KleinGordonEq}
\end{equation}
where $\Phi = \Phi(t, r, \psi, \theta, \phi)$ is the scalar field.
Due to the isometries of the backgrounds and the separability of the wave equation, we can expand the field in Fourier modes as:
\begin{equation}
    \Phi_{\omega p m} = \Psi(r)\,S(\theta)\,e^{i\left(\frac{p}{R_\psi} \psi+m \phi -\omega t\right)}.
    \label{eq:ScalarWaveProfile}
\end{equation}
where $m, p \in \mathbb{Z}$, since $\phi$ and $\psi$ are identified with periodicities $2\pi$ and $2\pi R_\psi$, respectively. A general scalar waveform is then expressed as
\begin{equation}
    \Phi = \sum_{p,m} \int d\omega\,\Phi_{\omega p m}.
\end{equation}
Substituting~\eqref{eq:ScalarWaveProfile} into the Klein–Gordon equation~\eqref{eq:KleinGordonEq} for the $\cW$-soliton metric~\eqref{eq:W_sol_metric} and the black string background~\eqref{eq:BlackString}, we find that both geometries yield the same angular equation:
\begin{equation}
    \frac{1}{\sin \theta} \partial_\theta \left(\sin \theta\,\partial_\theta S\right) \,-\, \left[\frac{1}{\sin^2 \theta} \left(m-\frac{p N}{2} (1+\cos \theta) \right)^2-\lambda\right] S =0,
    \label{eq:AngularEq}
\end{equation}
where $\lambda$ is the separation constant, and the charge $Q$ has been replaced by its quantized value using~\eqref{eq:QuantizationRelation}. The radial equation for the $\cW$-soliton becomes:
\begin{equation}
    \partial_r \left( r^2(2f_M-f_Q) \,\partial_r \Psi \right) \,-\, \left[\frac{r^2 f_M^3}{f_Q(2f_M-f_Q)} \frac{p^2}{R_\psi^2}-r^2f_Q \left(\omega+\frac{f_M}{2 r f_Q} N p \right)^2+\lambda \right] \Psi =0,
    \label{eq:RadialEqWSol}
\end{equation}
while for the black string it is given by:
\begin{equation}
    \partial_r \left( r^2f_- f_M \,\partial_r \Psi \right) \,-\, \left[\frac{r^2}{f_+} \frac{p^2}{R_\psi^2}-\frac{r^2f_-^2 f_+}{f_M} \left(\omega+\frac{1}{2 r f_+} N p \right)^2+\lambda \right] \Psi =0.
    \label{eq:RadialEqBS}
\end{equation}
Note that
$p/R_\psi$ plays the same role as $p_\psi$ in the geodesic analysis. It labels the tower of Kaluza–Klein modes and, from a four-dimensional perspective, induces an effective mass and charge under the Kaluza–Klein vector field. Specifically, the five-dimensional Lagrangian for a minimally coupled scalar field reduces, upon dimensional reduction along the fifth direction, to a four-dimensional Lagrangian describing a scalar field with effective mass and Kaluza–Klein charge:
\begin{equation}
    \partial_\mu \Phi \,\partial^\mu \Phi \,\to\, \left(\partial_a + i \frac{p}{R_\psi} \, g^{\psi}_{\,\,a} \right) \Phi \, \left(\partial^a + i \frac{p}{R_\psi} \, g^{\psi a}\right) \Phi - g^{\psi \psi} \frac{p^2}{R_\psi^2}\,\Phi^2,
\end{equation}
where the index $a$ runs over the four-dimensional coordinates.

While we will not discard the massive modes with $|p| \geq 1$ in the stability analysis, they correspond to scalar fields with large effective mass and Kaluza–Klein charge when $R_\psi$ is small compared to the object's size. Therefore, we will focus more closely on the phenomenologically relevant massless Kaluza–Klein mode with $p = 0$.

\subsection{Angular waveform}

The angular equation~\eqref{eq:AngularEq} corresponds to that of spin-weighted spherical harmonics with half-integer spin $s = -\frac{pN}{2} \in \tfrac{1}{2}\mathbb{Z}$~\cite{Goldberg:1966uu,SpinWHarm,Berti:2005gp,Dolan:2009kj,Brito:2015oca}. Regularity of the waveform at $\theta = 0$ and $\pi$ requires the separation constant to be positive and quantized as\footnote{Our quantization matches that of~\cite{Goldberg:1966uu,SpinWHarm,Berti:2005gp,Dolan:2009kj,Brito:2015oca}, namely $\lambda = \ell(\ell+1) - s^2$, with the shift $\ell_\text{theirs} = \ell_\text{ours} + s$ and $m_\text{theirs}=m_\text{ours}+s$. Our convention allows us to consider the quantum numbers $(\ell,m)$ to be always integers for arbitrary spin instead of having half-integer values for spin-half probes in \cite{Goldberg:1966uu,SpinWHarm,Berti:2005gp,Dolan:2009kj,Brito:2015oca}.}
\begin{equation}
    \lambda=\ell(\ell+1)+s(1+2\ell)= \ell(\ell+1)-\frac{pN}{2}(1+2\ell),\qquad \ell \in \mathbb{N},\qquad \ell \geq \frac{|p N|+p N}{2}, \qquad \ell \geq |m|.
    \label{eq:QuantizationLambda}
\end{equation}
After some algebra, we find that the regular angular waveform yields 
\begin{equation}
    S_{\ell,m,s}(\theta) = \mathcal{N}_{\ell,m,s}\,\left(\sin \frac{\theta}{2}\right)^{|m+2s|}\left(\cos \frac{\theta}{2}\right)^{|m|}\,P_{\ell-\frac{|m+2s|+|m|-2s}{2}}^{(|m+2s|,|m|)} (\cos \theta),
\end{equation}
where $P_n^{(\alpha,\beta)}(z)$ is the Jacobi polynomial\footnote{The Jacobi polynomial is given by $$P_n^{(\alpha, \beta)}(z) = \frac{(-1)^n}{2^n n!} (1 - z)^{-\alpha} (1 + z)^{-\beta} \frac{d^n}{dz^n} \left\{ (1 - z)^\alpha (1 + z)^\beta (1 - z^2)^n \right\}=\sum_{k=0}^{n} \binom{n + \alpha}{n - k} \binom{n + \beta}{k} \left( \frac{z - 1}{2} \right)^k \left( \frac{z + 1}{2} \right)^{n - k}.
$$} and ${\mathcal{N}_{\ell,m,s}}^2=\left(n+ \frac{\alpha+\beta+1}{2}\right) \frac{\left(n+\beta+1 \right)_\alpha}{\left(n+1 \right)_\alpha}$ is the normalization constant such that $S_{\ell,m,s}$ defines an orthonormal basis in $\ell$ and $m$. The $(\theta,\phi)$ profile can be associated to the spin-weighted spherical harmonics, ${}_{s} Y_{\ell',m'}(\theta,\phi)$, studied in~\cite{Goldberg:1966uu,Leaver:1985ax}: $S_{\ell,m,s}(\theta) e^{im\phi} = {}_{s} Y_{\ell+s,m+s}(\theta,\phi)$. 

For the massless Kaluza–Klein scalar mode with $p = 0$, or in the neutral case with $Q\propto N=0$, the angular profile reduces to the standard spherical harmonics, with $\lambda = \ell(\ell+1)$, indicating that these modes reproduce the standard four-dimensional physics of a massless scalar field. In contrast, higher modes with $p \geq 1$ do not support spherical harmonics. From a four-dimensional perspective, they behave as massive, charged probes with spin $s = -\frac{pN}{2}$.

\subsection{Schrödinger form and boundary conditions}

To analyze the radial wave equations, we recast them into a Schrödinger-like form by rescaling the waveform and introducing a tortoise-like radial coordinate ${r_*}$. For both the $\cW$-soliton and the black string, this transformation proceeds as follows:
\begin{align}
    &
    \text{$\cW$-soliton:}\hspace{1.9em} \Psi = \frac{\overline{\Psi}}{r \sqrt{f_M}},\qquad \frac{d{r_*}}{dr} = \frac{f_M}{2f_M-f_Q},\\
    &
    \text{Black string:}\quad \Psi = \frac{\overline{\Psi}}{r \sqrt{f_-}},\qquad \frac{d{r_*}}{dr} = \frac{1}{f_M},
\end{align}
These tortoise-like coordinates can be integrated explicitly, yielding:
\begin{align}
    &\label{eq:WSoltortoise}
    \text{$\cW$-soliton:}\hspace{1.9em} {r_*} = r-r_++\frac{r_+}{2} \log \left(\frac{r}{r_+}-1 \right)+\frac{r_-}{2}\log \left(\frac{r}{r_+}-1 \right),\\ 
    &\label{eq:BStortoise}
    \text{Black string:}\quad {r_*} = r-2M + 2M\log\left(\frac{r}{2M}-1 \right),
\end{align}
such that ${r_*} \to \infty$ corresponds to the asymptotic boundary ($r \to \infty$), while ${r_*} \to -\infty$ corresponds either to the smooth cap for the soliton or to the horizon for the black string, respectively.

\begin{equation}
    \frac{d^2 \overline{\Psi}}{{d{r_*}}^2} - V\,\overline{\Psi} \,=\,0,
\end{equation}
with effective potentials:
\begin{equation}
    \begin{split}
        V_\cW &= \frac{2f_M-f_Q}{f_M^2} \left[\frac{f_M^3}{f_Q(2f_M-f_Q)} \frac{p^2}{R_\psi^2}- f_Q\left(\omega+\frac{f_M}{2 r f_Q} N p \right)^2+\frac{\lambda}{r^2} + \frac{3f_Q(1+f_M^2)-2f_M(f_M^2-f_Q+3)}{4 r^2 f_M^2} \right],\\
        V_\text{BH} &= \frac{f_M}{f_- f_+} \frac{p^2}{R_\psi^2}-f_- f_+ \left(\omega+\frac{1}{2 r f_+} N p \right)^2+\frac{\lambda\,f_M}{r^2 f_-} + f_M \,\frac{2f_-(1+f_-)-f_M(1+3f_-^2)}{4 r^2 f_-^2},
    \end{split}
    \label{eq:PotentialsDef}
\end{equation}
where $V_\cW$ and $V_\text{BH}$ denote the scalar potentials for the $\cW$-soliton and the black string, respectively.\footnote{
For $Q=0$ or $p=0$, using a different tortoise-like coordinate and field redefinition, it is also possible to recast the equation in a more canonical form as $d^2\tilde{\Psi}/dr_*^2+(\omega^2-{V_{\rm eff}})\tilde{\Psi}=0$, where ${V_{\rm eff}}$ is frequency independent, at the cost of introducing transformations and effective potentials which are singular at the inner boundary.}
Owing to the isometries of the background, the radial equation does not depend on the azimuthal number $m$, so the corresponding QNM spectrum will be degenerate in $m$.

At large distances, both potentials asymptote to $\frac{p^2}{R_\psi^2} - \omega^2$. For the $\cW$-soliton, due to the absence of dissipation at the inner boundary, regular scalar modes are either trapped normal modes when $\frac{p}{R_\psi} > |\omega|$, or quasinormal (dissipative) modes when $\frac{p}{R_\psi} < |\omega|$. As previously discussed, $\frac{p}{R_\psi}$ acts as an effective mass, so the trapped modes correspond to massive scalar excitations confined within the geometry. 
The boundary condition at infinity corresponds to outgoing waves for QNMs with $\frac{p}{R_\psi} < |\omega|$ and decaying waves for normal modes with $\frac{p}{R_\psi} > |\omega|$:
\begin{equation}
    \Psi(r) e^{-i\omega t} \underset{{r_*}\to\infty}{\propto} \frac{1}{{r_*}}e^{i \left(\sqrt{\omega^2-\frac{p^2}{R_\psi^2}} \,{r_*}-\omega t\right)}.
    \label{eq:BCasymp}
\end{equation}

At the black string horizon (${r_*} \to -\infty$), the potential tends to a finite negative value $-{\omega'}^2$, with,
\begin{equation}
 \omega' = \frac{\sqrt{3M-\sqrt{M^2+4Q^2}}\left((M+\sqrt{M^2+4Q^2})\,\omega+\frac{pN}{2}\right)}{2M\sqrt{M+\sqrt{M^2+4Q^2}}},
 \label{eq:omegaprime}
\end{equation}
and regular solutions correspond to ingoing modes:
\begin{equation}
    \Psi(r) e^{-i\omega t} \underset{{r_*}\to-\infty}{\propto} e^{-i \left(\omega' {r_*}+\omega t\right)}.
    \label{eq:BCinnerBS}
\end{equation}

For the $\cW$-soliton, the potential near the smooth cap (${r_*} \to -\infty$) approaches either a finite positive value if $p \neq 0$, or zero if $p = 0$. Regularity requires Neumann boundary conditions in terms of the local radial coordinate $\rho^2 = 4(r - r_+) \sim 4r_+ e^{\frac{{r_*}}{r_+}}$, leading to:
\begin{equation}
    e^{-\frac{{r_*}}{2r_+}}\,\partial_{{r_*}} \Psi \underset{{r_*}\to-\infty}{=}0,
    \label{eq:BCbolt}
\end{equation}
which implies that $\Psi$ must converge to a constant faster than $e^{\frac{{r_*}}{2r_+}}$ at ${r_*}\to -\infty$.

\subsection{Scalar stability of $\cW$-soliton} \label{sec:stability}

To assess the stability of $\cW$-solitons under scalar perturbations, we search for critical static modes with $\omega = 0$, (see, e.g., Refs.~\cite{Reall:2001ag,Gubser:2001ac,Gubser:2002yi,Miyamoto:2007mh,Gregory:1993vy} for similar studies in other contexts). For $\omega = 0$, the potential defined in~\eqref{eq:PotentialsDef} is everywhere positive for all values of $p$, $\lambda$, $M$, and $Q$ within the range specified in~\eqref{eq:ValidityBound}. Consequently, no solution can remain regular at both boundaries, ${r_*}\to \infty$ and $-\infty$; any mode must diverge at one of them. Hence, $\cW$-solitons do not admit critical static modes. In some upcoming sections, by performing an extensive numerical analysis, we will confirm this results and extend it throughout the entire parameter space, also excluding unstable modes whose threshold does not occur at a critical static mode.


\subsection{WKB approximation of the QNMs}

The WKB approximation provides an effective method for approximating solutions to the Schrödinger equation when the potential varies slowly and the imaginary part of $\omega$ is negligible compared to the real part. This corresponds to the regime in which the modes are slowly damped relative to the timescale of their oscillations. In our context, these conditions are typically satisfied when
\begin{equation}
    \ell \gg 1,\qquad \omega_R \gg \omega_I,\qquad \omega \equiv \omega_R + i\,\omega_I.
\end{equation}
We also search for QNMs satisfying $\frac{p}{R_\psi} < |\omega|$, such that the wave remains oscillatory at large distances, in agreement with the boundary condition~\eqref{eq:BCasymp}.

\subsubsection{Slowly-damped QNMs of the $\cW$-soliton}

We begin by applying WKB analysis to scalar QNMs of the $\cW$-soliton. After a detailed examination of the potential~\eqref{eq:PotentialsDef}, we find that at most one root appears in the domain $-\infty < {r_*} < \infty$, and the potential is either entirely negative or becomes negative at large ${r_*}$ and positive at small ${r_*}$. When the potential is everywhere negative, the WKB approximation does not yield any regular wave solutions that satisfy both boundary conditions.\footnote{If the potential does not change sign, the wave is approximated by a linear combination of $\overline{\Psi}_\pm$ across the entire range of ${r_*}$. In such cases, it is impossible to satisfy both the outgoing boundary condition at large distances~\eqref{eq:BCasymp} and regularity at the bolt~\eqref{eq:BCbolt}.}

\begin{figure}[t]
  \centering
\includegraphics[width=0.72\textwidth]{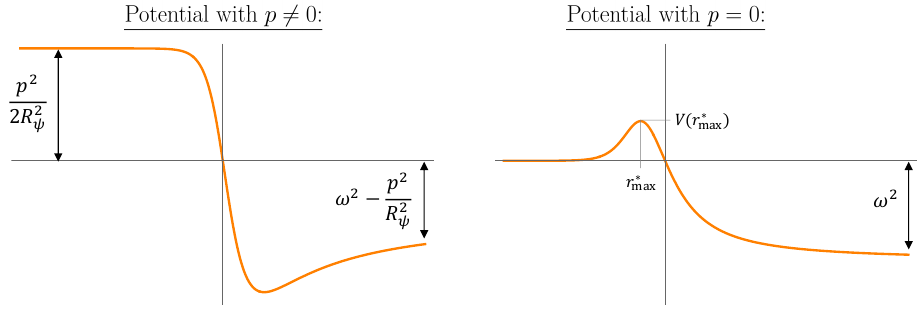}
  \caption{Scalar potential of the $\cW$-soliton for $p = 0$ and $p \neq 0$, in regimes where the potential changes sign. The plots show $V_\cW({r_*})$ from~\eqref{eq:PotentialsDef} for $k = 100$, $N = 30$, $\ell = 10$, and $(\omega R_\psi, p) = (1.03, 1)$ or $(1/3, 0)$.}
  \label{fig:PotentialWSol}
\end{figure}

We therefore focus on cases where the potential changes sign. Two qualitatively distinct types of potentials arise, as shown in Fig.~\ref{fig:PotentialWSol}: when $p \neq 0$, the potential exhibits a centrifugal barrier near the cap; when $p = 0$, no such barrier is present.

For $p \neq 0$, the wavefunction is approximated by WKB solutions on either side of the turning point, one in the positive-potential region and one in the negative-potential region, connected by a standard Airy function matching at the turning point. In each region, the wave approximates as a linear combination of two ``$\pm$'' branches:
\begin{equation}
    \overline{\Psi}_\pm =  |V(r^*)|^{-\tfrac{1}{4}} 
\exp\!\left[ \pm \int^{r^*} \sqrt{V(r^*)}\, dr^* \right].\label{eq:WKBfunction}
\end{equation}

The asymptotic boundary condition~\eqref{eq:BCasymp} requires selecting the ``$+$'' branch in the negative-potential region. However, the Airy matching conditions at the turning point enforce a linear combination of ``$+$'' and ``$-$'' branches in the positive-potential region. The ``$-$'' branch is necessarily singular at the bubble locus, and thus the wave cannot be regular. As a result, no slowly damped QNMs are captured by the WKB approximation when $p\neq 0$. This does not exclude the existence of QNMs altogether, but any such modes must have a non-negligible imaginary component of $\omega$, implying they are rapidly damped.

For $p = 0$, the same conclusion applies when the maximum of the potential is large. However, when the maximum is small (as shown in the right plot of Fig.~\ref{fig:PotentialWSol}), the wavefunction in the negative-potential region can be described by the WKB approximation~\eqref{eq:WKBfunction}, and directly matched to an exact solution of the Schrödinger equation in the small bump potential near the origin. As shown in~\cite{Heidmann:2023ojf}, regular, slowly damped QNMs exist in this regime if the following quantization condition is satisfied for an overtone number $n\geq 0$:\footnote{Our convention differs slightly from~\cite{Heidmann:2023ojf}, with an overall minus sign on the right-hand side of the quantization condition. This arises from a different choice in the time dependence of the wave, $e^{\pm i \omega t}$.}
\begin{equation}
    i \,\frac{\sqrt{2}\,V_\cW({r_*}_\text{max})}{\sqrt{|V_\cW''({r_*}_\text{max})|}} = -n-\frac{1}{2},\qquad n\in \mathbb{Z},\quad n\geq 0.
\end{equation}
Assuming large $\ell$ and $\omega = \mathcal{O}(\ell)$, the potential becomes $V_\cW \approx -R(r)$ with $E=-\omega$ and $C=\lambda$, where $R(r)$ is the geodesic potential~\eqref{eq:Theta-R}, such that the properties of the potential around the maximum are intimately connected to the properties of the photon orbit. The QNM frequencies are well approximated by~\cite{Ferrari:1984zz,Cardoso:2008bp,Heidmann:2023ojf}:
\begin{equation}
    \omega_{\cW;\ell,n} \,=\, \Omega_\cW \,\left(\ell+\frac{1}{2} \right) - i\,\lambda_\cW \,\left(n+\frac{1}{2} \right), \qquad n\in \mathbb{Z},\quad n\geq 0,
\end{equation}
where $\Omega_\cW$ is the angular velocity of null geodesics at the photon sphere, and $\lambda_\cW$ is the Lyapunov exponent as defined in~\eqref{eq:PhotonOrbitPropWSol}. The WKB approximation remains valid as long as $\omega_R \gg \omega_I$, which translates to $\frac{2\ell+1}{2n+1} \gg \sqrt{2}$ since for the $\cW$-soliton, $\lambda_\cW/\Omega_\cW = \sqrt{2}$.

\subsubsection{Slowly-damped QNMs of the black string}

We now apply the WKB approximation to derive the spectrum of slowly damped scalar QNMs of the black string solution. The black string potential~\eqref{eq:PotentialsDef} asymptotes to negative values: $-\omega'^2$ ~\eqref{eq:omegaprime} as ${r_*} \to -\infty$, and $-\omega^2 + \frac{p^2}{R_\psi^2}$ as ${r_*} \to +\infty$. In between, the potential can develop a positive ``bump'' region. This characteristic profile is typical of static black hole spacetimes and has been studied in simpler geometries such as Schwarzschild~\cite{Schutz:1985km,Iyer:1986np}.

Slowly damped modes captured by the WKB approximation arise when the positive-potential region is narrow and well approximated by a parabola. Regularity of the QNMs at both the horizon and infinity leads to the following quantization condition~\cite{Iyer:1986np}:
\begin{equation}
    i \,\frac{V_\text{BH}({r_*}_\text{max})}{\sqrt{2|V_\text{BH}''({r_*}_\text{max})|}} = -n-\frac{1}{2},\qquad n\in \mathbb{Z},\quad n\geq 0.
\end{equation}
where ${r_*}_\text{max}$ denotes the location of the potential maximum.

For general $p \neq 0$, there is no analytic simplification to determine ${r_*}_\text{max}$ and invert the quantization condition to obtain the QNM spectrum $\omega_{\text{BH};\ell,n}$. However, in the special case $p = 0$, and assuming large $\ell$ with $\omega = \mathcal{O}(\ell)$, we find that $V_\text{BH} \approx -\widetilde{R}(r)$ with $E = -\omega$ and $C = \lambda$ where $\widetilde{R}(r)$ is the geodesic potential~\eqref{eq:RdefBS}. The quantization relation then yields the well-known WKB QNM spectrum~\cite{Cardoso:2008bp}, expressed in terms of the photon sphere characteristics, similarly to the $\cW$-soliton:
\begin{equation}
    \omega_{\text{BH};\ell,n} \,=\, \Omega_\text{BH} \,\left(\ell+\frac{1}{2} \right) - i\,\lambda_\text{BH} \,\left(n+\frac{1}{2} \right), \qquad n\in \mathbb{Z},\quad n\geq 0,
\end{equation}
where $\Omega_\text{BH}$ is the angular velocity of photons at the photon sphere, and $\lambda_\text{BH}$ is the Lyapunov exponent defined in~\eqref{eq:PhotonOrbitPropBS}. The WKB approximation is valid in the regime $\omega_R \gg \omega_I$, which corresponds to $\frac{2\ell+1}{2n+1} \gg 1$. For the black string, the ratio $\lambda_\text{BH}/\Omega_\text{BH}$ lies in the range $\frac{1}{\sqrt{2}} \leq \lambda_\text{BH}/\Omega_\text{BH} \leq 1$.

\subsubsection{Comparison}

\begin{figure}[t]
  \centering
\includegraphics[width=0.45\textwidth]{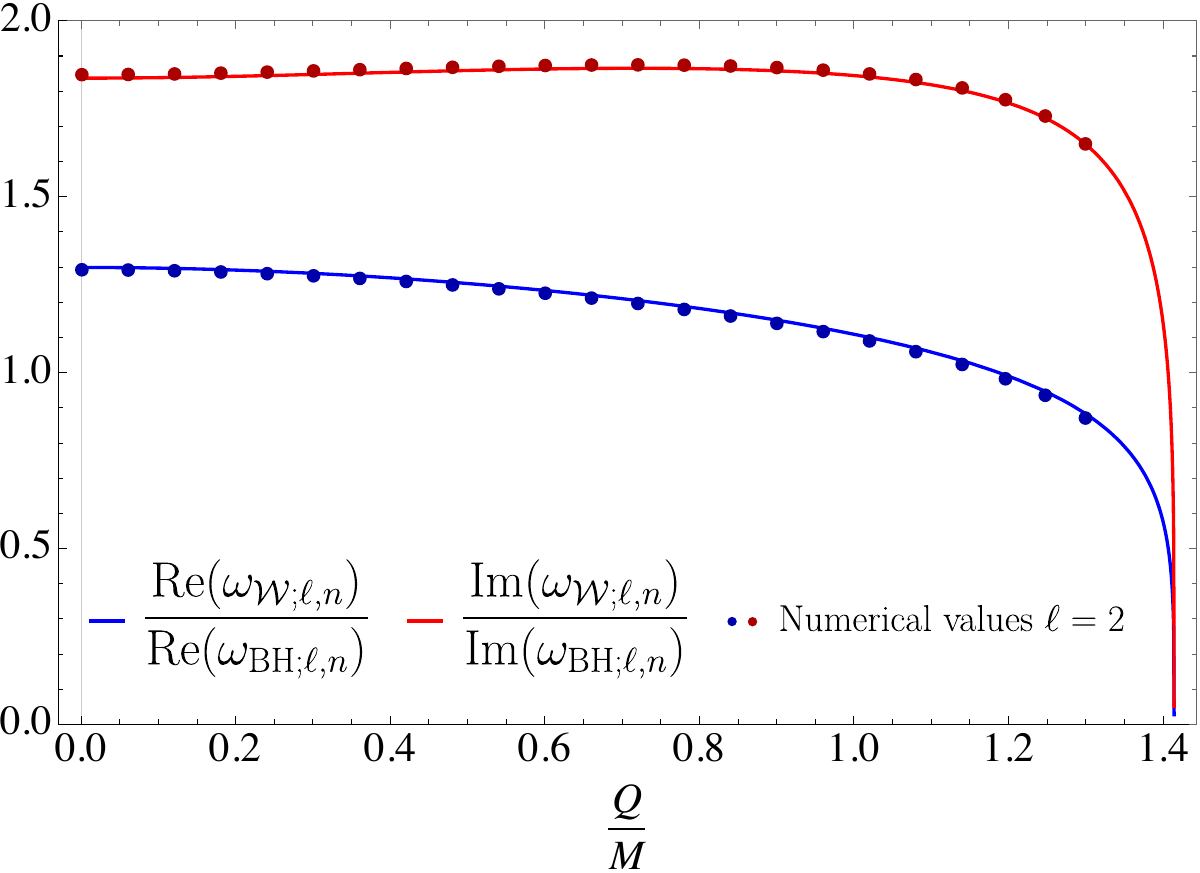}
  \caption{Ratios of the QNM frequencies for the $\cW$-soliton and the black string. In the WKB approximation (solid lines), these ratios depend only on $Q/M$ and are independent of $(M,\ell,n)$. The dots are the numerical values from the direct-integration method at $\ell=2$ (up to $Q/M=1.3$), indicating an excellent agreement with WKB values, even for small $\ell$.
  } 
  \label{fig:QNMWKB}
\end{figure}

The WKB analysis of the QNMs of an effectively massless scalar field ($p=0$) reveals that the slowly damped modes in both the $\cW$-soliton and the black string are governed by the properties of their photon spheres, specifically, the angular velocities at the unstable circular orbit and the associated Lyapunov exponents.

Since the photon spheres of the $\cW$-soliton and the black string differ at $\mathcal{O}(1)$, their QNM frequencies do not coincide when $\ell$, $n$, $M$, and $Q$ are held fixed. In Fig.~\ref{fig:QNMWKB}, we plot the ratios of the real and imaginary parts of the QNM frequencies as functions of $Q/M$. For $|Q|/M \leq 1.19$, the $\cW$-soliton exhibits QNMs with larger frequencies and greater damping rates, indicating that its ringdown signal oscillates more rapidly and decays more quickly than that of the black string far from extremality. The WKB results (solid lines Fig.~\ref{fig:QNMWKB}) are in excellent agreement with the ``true'' QNM spectrum (dots in Fig.~\ref{fig:QNMWKB}), computed numerically as discussed in the next section.

Furthermore, the absence of a stable photon sphere in the $\cW$-soliton geometry precludes the existence of long-lived QNMs associated with late-time echoes~\cite{Heidmann:2023ojf,Bianchi:2023sfs,Dima:2024cok,Bena:2024hoh,Dima:2025zot}. This implies that the scalar response of the $\cW$-soliton and the black string should slightly differ during the early and intermediate ringdown, but become indistinguishable, with no echoes present and the same universal power-law tail behavior at late time~\cite{Rosato:2025rtr}. According to the WKB approximation, the ringdown signal of each spherical harmonic component ($\ell$ mode) of the $\cW$-soliton should exhibit a shorter oscillation timescale and a faster decay compared to the black string. In the next sections, we will extend this result with numerical analysis both in the frequency and in the time domain.

\subsection{Numerical QNM spectrum} \label{sec:fullQNM}

In this section, we present our numerical derivation of the full QNM spectrum for an effectively massless scalar field ($p=0$), in both the neutral $\cW$-soliton and the Schwarzschild black hole.

We computed the QNM frequencies numerically using both the continued fraction method and a direct-integration shooting method (see e.g.\cite{Ferrari:2007rc,Rosa:2011my,Pani:2013pma,Pani:2012bp,Leaver:1985ax} for details), finding excellent agreement. Table~\ref{tab:test_scal_p0_QNMs} presents the results for the fundamental ($n=0$) QNM, normalized by the mass $M$, for several values of $\ell$, together with the corresponding WKB values.

\begin{table*}[h!]
    \centering
    \begin{tabular}{|c||c|c||c|c|}
    \hline
    $\ell$ & \multicolumn{2}{c||}{Neutral $\cW$-soliton} & \multicolumn{2}{c|}{Schwarzschild BH} \\
    \hline
    \hline
      & continued fraction & WKB   &continued fraction & WKB  \\
    \hline
    $0$ & $0.13669 - \I 0.20545$ & $0.125-\I 0.17678$ & $0.11046 - \I 0.10490$ & $0.09623 - \I 0.09623$\\
    \hline
    $1$ & $0.37542 - \I 0.18194$ & $0.375 -\I 0.17678$ & $0.29294 - \I 0.09766$ & $0.28868 - \I 0.09623$ \\
    \hline
    $2$ & $0.62508 - \I 0.17872$ & $0.625 -\I 0.17678$ & $0.48364 - \I 0.09676$ & $0.48113 - \I 0.09623$ \\
    \hline
    $3$ & $0.87503 - \I 0.17778$ & $0.875 -\I 0.17678$  & $0.67537 - \I 0.09650$ & $0.67358 - \I 0.09623$ \\
    \hline
    $5$ & $1.37501 - \I 0.17719$ & $1.375 -\I 0.17678$ & $1.05961 - \I 0.09634$ & $1.05848 - \I 0.09623$\\
    \hline
    $10$ & $2.62500 - \I 0.17689$ & $2.625 -\I 0.17678$ & $2.02132 - \I 0.09626$ & $2.0207 - \I 0.09623$\\
    \hline
    $30$ & $7.62500 - \I 0.17679$ & $7.625 -\I 0.17678$ & $5.86993 - \I 0.09623$ & $5.8697- \I 0.09623$ \\
    \hline
    \end{tabular}
    \caption{Fundamental ($n=0$) QNMs of an effectively massless scalar field ($p=0$) in the neutral ($Q=0$) $\mathcal{W}$-soliton and Schwarzschild black hole backgrounds, for various harmonic numbers $\ell$. We show the dimensionless quantity $\omega M$ from the continued fraction method and the WKB values.
    }
    \label{tab:test_scal_p0_QNMs}
\end{table*}

Strikingly, the WKB approximation already achieves an error below 1\% at $\ell=1$ for both the soliton and the black hole, underscoring its excellent accuracy in capturing the slowly damped modes of the spectrum, even for small $\ell$.

The remarkable agreement between the WKB approximation and the numerical results observed for neutral configurations persists for the charged solutions, even at low $\ell$. This matching is illustrated in Fig.~\ref{fig:QNMWKB} where we have inserted the numerical values obtained on top of the WKB result. For practical reasons, the numerical data terminate around $Q/M \sim 1.3$, rather than extending to extremality, because the numerical cost increases substantially as extremality is approached, with the QNM frequencies becoming progressively larger.\\

While the WKB approximation provides reliable predictions for the fundamental QNM frequencies, it fails to capture modes where the imaginary part of the frequency is comparable to or larger than the real part: the {highly damped modes}. Table~\ref{tab:test_scal_p0_QNMs_wOvertones} presents the numerical values of the first few highly damped modes for $\ell=1$ and $\ell=5$ in both the Schwarzschild black hole and the neutral $\cW$-soliton backgrounds.

\begin{table*}[h!]
    \centering
    \begin{tabular}{|c||c|c||c|c|}
    \hline
    $n$ & \multicolumn{2}{c||}{Neutral $\cW$-soliton} & \multicolumn{2}{c|}{Schwarzschild BH} \\
    \hline
    \hline
      & $\ell=1$ & $\ell=5$    &$\ell=1$ & $\ell=5$   \\
    \hline
    $0$ & $0.37542 - \I 0.18194$ & $1.37501 - \I 0.17719$ & $0.29294 - \I 0.09766$ & $1.05961 - \I 0.09634$ \\
    \hline
    $1$ & $0.28371 - \I 0.60411$ & $1.34139 - \I 0.53626$ & $0.26445 - \I 0.30626$ & $1.05004 - \I 0.29015$ \\
    \hline
    $2$ & $0.20907 - \I 1.10779$ & $1.27631 - \I 0.90978$ & $0.22954 - \I 0.54013$ & $1.03150 - \I 0.48735$ \\
    \hline
    $3$ & $0.17171 - \I 1.62472 $ & $1.18556 - \I 1.30773$ & $0.20326 - \I 0.78830$ & $1.00520 - \I 0.68993$ \\
    \hline
    $4$ & $0.15050 - \I 2.13882 $ & $1.08008 - \I 1.73772$ & $0.18511 - \I 1.04076$ & $0.97297 - \I 0.89948$ \\
    \hline
    \end{tabular}
    \caption{Overtone QNMs of an effectively massless scalar field ($p=0$) in the neutral ($Q=0$) $\mathcal{W}$-soliton and Schwarzschild black hole for different overtones $n$ and for $\ell=1$ and $\ell=5$.}
    \label{tab:test_scal_p0_QNMs_wOvertones}
\end{table*}

We observe that the dependence of the highly damped frequencies on the overtone number $n$ is broadly similar for the Schwarzschild black hole and the neutral $\cW$-soliton: the imaginary part grows and the real part decays with $n$ in much the same way. Thus, although the fundamental frequency ($n=0$) differs significantly, the higher-overtone towers as a function of $n$ display very similar characteristics in both cases. This indicates that the time-domain ringdown response of the soliton and the Schwarzschild black hole should share a similar structure, differing primarily in the fundamental oscillation frequency and the overall damping envelope. We analyze this point in detail in the next section.

\subsection{Time-domain ringdown} \label{sec:timedomain}

In this section, we present the linear response of the black string and the $\cW$-soliton to massless scalar perturbations in the time domain. This response is associated to the ringdown signal of the backgrounds under scalar perturbation. We will have a particular interest in the neutral configurations, the Schwarzschild black hole and the neutral $\cW$-soliton. We first outline the numerical implementation and then discuss the simulations, analyzing how the soliton’s behavior differs from that of the black hole and how the results compare with the frequency-domain analysis of QNM spectra from the previous sections.

\subsubsection{Numerical method}

We study the time evolution of test scalar perturbations by numerically solving the radial equations, Eqs.~\eqref{eq:RadialEqWSol} and~\eqref{eq:RadialEqBS} but keeping an arbitrary time evolution\footnote{This can be achieved by considering $\Psi=\Psi(t,r)$ and replacing $\omega \to i \partial_t$ in
Eqs.~\eqref{eq:RadialEqWSol} and~\eqref{eq:RadialEqBS}.
} instead of $\sim e^{-i\omega t}$, with suitably chosen initial data. For the neutral $\cW$-soliton, all quantities are expressed in terms of the radial coordinate introduced in Section~\ref{sec:WSolGen}, $R(r)=\sqrt{g_{\theta \theta}} = r\sqrt{f_M}$. For the black hole, the equation for the test field is written using the tortoise coordinate, \eqref{eq:BStortoise}, allowing a convenient treatment of the near-horizon region.

We solve the scalar equations with a custom PDE code implemented in \Julia, based on the method of lines: time integration is performed with a fourth-order Runge–Kutta algorithm, and spatial derivatives are approximated by standard fourth-order finite-difference stencils. Boundary conditions are imposed via fourth-order finite-difference schemes. For the black string, we impose ingoing/outgoing radiative boundary conditions, Eqs.~\eqref{eq:BCinnerBS} and~\eqref{eq:BCasymp}, at the inner and outer boundaries, respectively. For the $\cW$-soliton, we enforce a Neumann condition at the smooth cap \eqref{eq:BCbolt}, identified via a Fuchsian analysis of the radial equation near the bubble locus $r=r_+$.

The initial data are chosen as a Gaussian-shaped perturbation along the radial direction, typically centered at $R_0 = 100 M$ at $t=0$, and given by a specific spherical harmonic $Y_{\ell 0}(\theta,\phi)$ along the angular directions, so that each $\ell$ mode can be excited individually.
To ensure that the source excites the QNMs,\footnote{If the source does not excite the QNMs, then the response will display a universal waveform that does not depend on the characteristics of the geometries.} the Fourier transform of the initial profile must overlap with the relevant QNM frequency range. Accordingly, we adjust the Gaussian width so that the perturbation efficiently excites modes in the desired frequency band. In particular, since the real part of the QNM frequency scales approximately linearly with $\ell$, exciting higher-$\ell$ modes requires narrower initial profiles.

The ringdown responses are then obtained by extracting the values of the test field at a sufficiently large extraction radius, typically $R_\text{ext} = 100 M$.

\subsubsection{Analysis of scalar ringdowns}

\begin{figure}[th]
  \centering
  \includegraphics[width=0.6\textwidth]{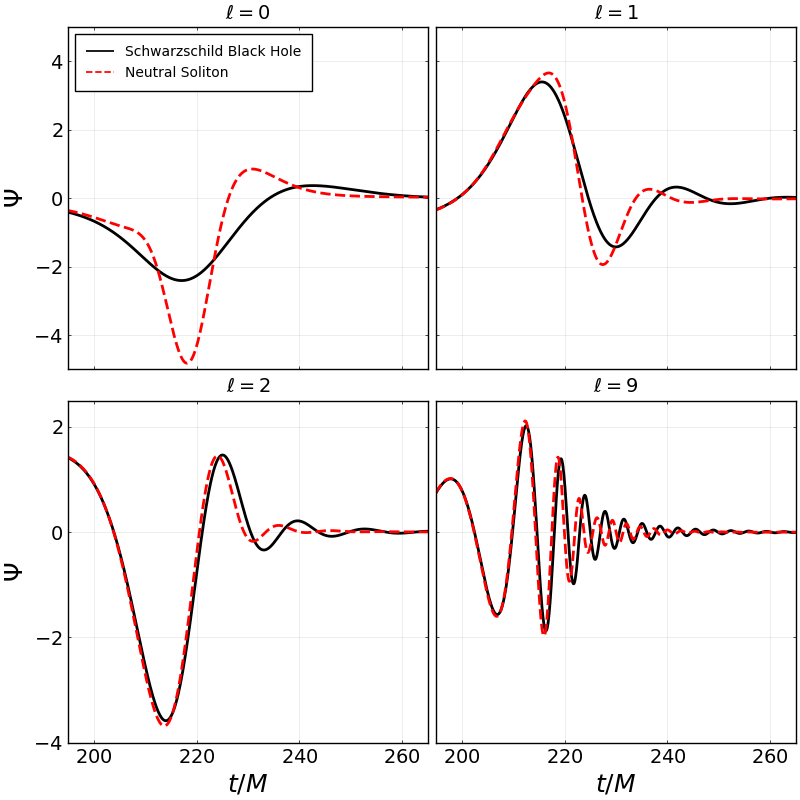}
  \caption{Ringdown responses of a test scalar perturbation $\Psi$ as a function of time and evaluated at $R_\text{ext}=100M$ for the Schwarzschild black hole and the neutral $\cW$-soliton. We considered four illustrative spherical harmonics $\ell=\{0\,,1\,,2\,,9\}$.}
\label{fig:ringdown_comparison_linear}
\end{figure}

We first apply the above procedure to the neutral configurations and obtain the linear scalar ringdown responses of the neutral $\cW$-soliton and the Schwarzschild black hole for illustrative spherical harmonics $\ell=0,1,2$, and $9$.
The corresponding waveforms are shown in Fig.~\ref{fig:ringdown_comparison_linear}.

For $\ell=1,2$, and $9$, the soliton and Schwarzschild signals display a similar structure: (i) no echoes are present in the soliton response, (ii) the very early part of the signal closely matches that of the black hole, and (iii) in the intermediate signal the soliton shows a slightly higher oscillation frequency and a faster damping rate.
These features agree with the frequency-domain analysis. Point (i) follows from the absence of long-lived modes in the soliton (no stable light rings), point (ii) reflects the similar spectrum of highly damped, low-frequency modes,\footnote{Note that the very early part of the signal is largely determined by the prompt response, which depends on the initial data. The prompt response is similar for both the soliton and the black hole, although in the former case also includes the first reflection of the initial data at the inner boundary.} and point (iii) is explained by the properties of the fundamental QNM, whose real and imaginary parts are both slightly larger for the soliton than for Schwarzschild (Table~\ref{tab:test_scal_p0_QNMs}), implying faster oscillation and decay in the time domain.

The $\ell=0$ ringdown behaves somewhat differently: the soliton response is much stronger than that of Schwarzschild. This significant enhancement is due to the spherical symmetry of the perturbation that suits the spherical symmetry of the background and its strong reflectivity at the bubble locus. Despite this, we retrieve the absence of long-lived modes and a larger fundamental frequency and damping for the soliton with respect to Schwarzschild. \\

\begin{figure}[th]
  \centering
\includegraphics[width=0.9\textwidth]{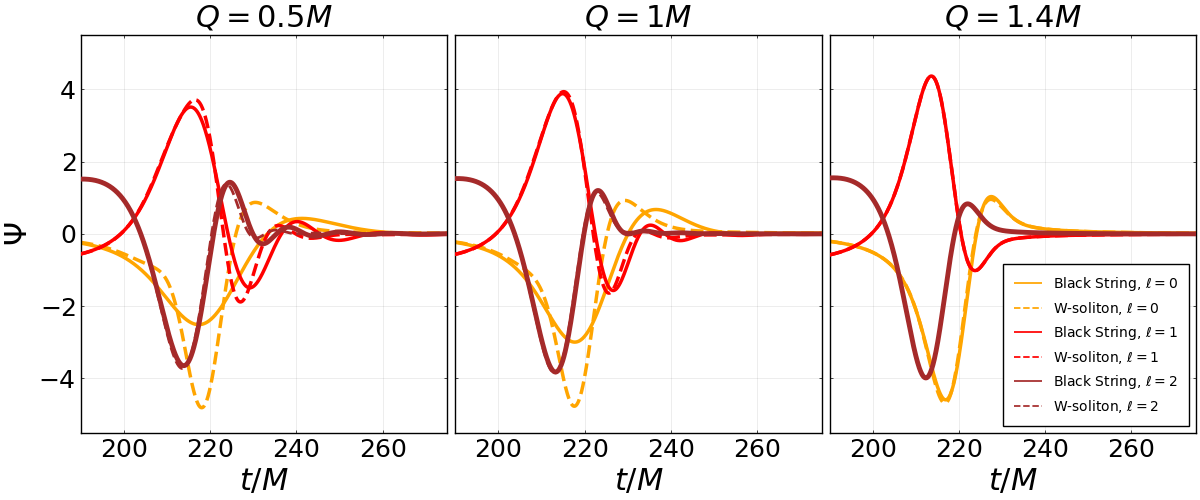}
  \caption{Ringdown responses of a test scalar perturbation $\Psi$ as a function of time and evaluated at $R_\text{ext}=100M$ for black strings and $\cW$-soliton with charge $Q=\{0.5\,,1\,,1.4\}M$ (respectively, left, middle and right panels). We considered three harmonics $\ell=\{0\,,1\,,2\}$ (respectively, orange, red and brown curves).}
\label{fig:ringdown_comparison_linear_wsol}
\end{figure}

We then extend the analysis to charged configurations. Figure~\ref{fig:ringdown_comparison_linear_wsol} shows representative ringdown signals for $\ell=0,1,2$ and three charge values approaching the extremality bound ($Q=\sqrt{2}M$).

As expected, when the charge approaches extremality, the two geometries become nearly indistinguishable, and so do their ringdown signals. Otherwise, the behavior closely parallels the neutral case: the soliton’s response is qualitatively similar to that of the charged string, but with different oscillation and damping times in the intermediate signal, as suggested by the WKB analysis in Fig.~\ref{fig:QNMWKB}. Note, however, that the WKB result indicates significant differences between the QNM frequencies of the two backgrounds even near extremality. For example, at $Q=1.4M$ the fundamental QNM of the black string is expected to oscillate about $1.7$ times faster and be $1.1$ times less damped than that of the soliton. This is not what is observed in Fig.~\ref{fig:ringdown_comparison_linear_wsol}. The reason is that the fundamental frequencies are different but very high close to extremality, and therefore a very narrow initial Gaussian perturbation is required to excite these modes.\footnote{Close to extremality, the real and imaginary parts of the QNMs are set by the angular velocity and Lyapunov exponent of photons at the photon sphere according to the WKB analysis which diverge at extremality (see Fig.~\ref{fig:PhotonSphereProp}).} Close to extremality, both geometries approach a point-like geometry (a Gibbons–Hawking center; see Sec.~\ref{sec:WSolGen}), so only perturbations with very short wavelengths (large frequency) can probe the geometry; otherwise the response is determined by the prompt response which depends on the initial data as obtained in Fig.\ref{fig:ringdown_comparison_linear_wsol}. \\

\begin{figure}[th]
  \centering
\includegraphics[width=0.8\textwidth]{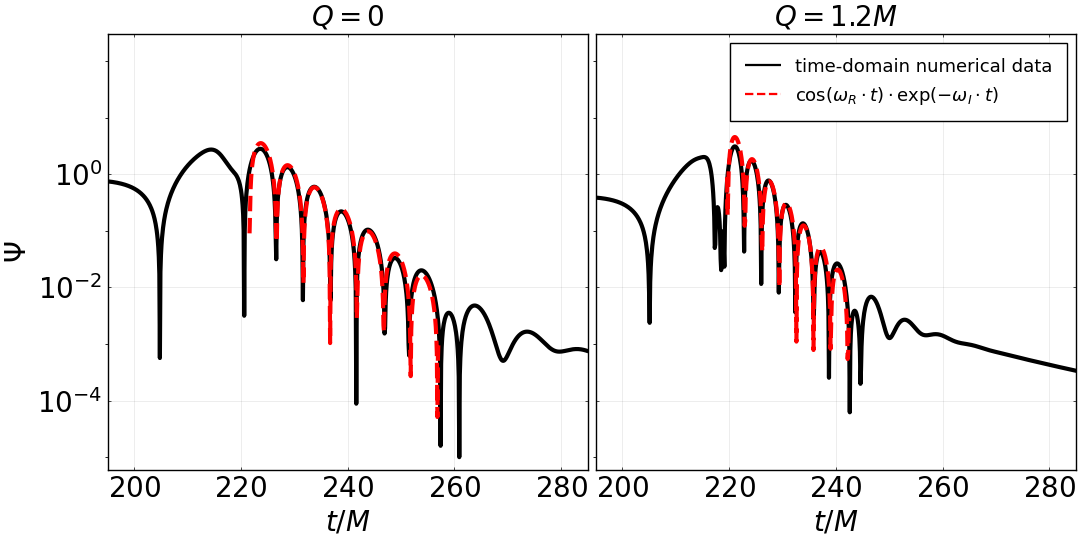}
  \caption{Comparison of the time-domain response with a single-mode damped sinusoid using the QNM estimate, for $\cW$-solitons with $Q=0$ (left) and $Q=1.2M$ (right) for test scalar perturbations with $\ell=2$. Note that the soliton ringdowns in this figure were excited via initial data with a narrower Gaussian radial profile than the one  signals in Fig.~\ref{fig:ringdown_comparison_linear}.
  }
\label{fig:qnm_tvsf}
\end{figure}

Finally, we aimed to determine the extent to which the ringdown responses of the soliton are governed by the fundamental QNM frequency.
Figure~\ref{fig:qnm_tvsf} compares the ringdown (shown on a logarithmic scale to mitigate the effect of strong damping) with a single damped sinusoid constructed solely from the fundamental QNM of the $\cW$-soliton.
We focus on two representative cases: an $\ell=2$ perturbation of a neutral soliton (left panel) and of a $Q=1.2 M$ soliton (right panel). For these runs, we used a slightly different Gaussian initial data, producing a slightly different prompt response at early times compared with the simulations in Figs.~\ref{fig:ringdown_comparison_linear_wsol} and \ref{fig:ringdown_comparison_linear}.
The first two oscillations clearly deviate from a single sinusoid, reflecting the fact that the very early signal is determined by the prompt response and the superposition of several highly damped modes.
Beyond this initial stage, however, the fundamental mode alone accurately reproduces the intermediate part of the ringdown, indicating that higher overtones are strongly suppressed by their short damping times (see Table~\ref{tab:test_scal_p0_QNMs_wOvertones}).
This confirms that, after a brief transient, the signal is dominated by the fundamental mode.

\subsection{Numerical evidence for scalar stability}

In Section~\ref{sec:stability}, we presented an analytic argument proving the absence of unstable modes with a static-mode threshold for the $\cW$-solitons. This suggests that $\cW$-solitons are stable under arbitrary scalar perturbations.

Here, we performed numerical simulations to support this result and extend it to generic scalar perturbations, even if the threshold of the putative instability does not occur at $\omega_R=0$. 
We searched for growing modes across various background charges $Q$ and harmonics $\ell$, and also considering a nonvanishing Kaluza–Klein momentum $p$.
The latter was not included in the previous simulations because, from a four-dimensional perspective, it corresponds to perturbations that are both highly massive and charged (see Section~\ref{sec:ScalWaveGen}).
We therefore carried out extensive simulations with $p\neq 0$ and found no evidence for linear scalar instability for both the solitons \emph{and} the black strings: scalar perturbations show no time growth.

When $p\neq0$ in the $\cW$-soliton, however, we did observe the existence of \emph{bound} states, i.e. purely normal modes, trapped between the inner reflective boundary and the effective mass potential.
This behavior is expected, since waves with $p\neq 0$ behave as massive modes from the four-dimensional point of view, and smooth horizonless geometries can support such massive probes ``orbiting'' the geometry without dissipation.

\section{Discussion and Conclusions}
In this work, we have analyzed the geometric properties and phenomenology of $\cW$-solitons — a class of smooth, horizonless solutions in five-dimensional minimal supergravity — as prototypes of black hole microstates. These solitons share the same mass and charge as four-dimensional charged black holes and reproduce several of their key features, while avoiding horizons and resolving the would-be inner singularity through a smooth topological cap supported by electromagnetic fluxes. Remarkably, they achieve this without introducing any additional tunable parameters beyond the mass (and possibly the charge). As a result, unlike many phenomenological models of exotic compact objects, the solution has a sharp signature that can, in principle, be directly tested by observations.
From a four-dimensional perspective, they correspond to (singular) charged geometries with flux, a massless axion, and the usual dilaton arising from Kaluza–Klein reduction. Remarkably, $\cW$-solitons also admit a neutral, asymptotically flat limit that asymptotically reproduces the Schwarzschild metric, providing an analytically tractable and astrophysically relevant configuration.

We have derived the response of $\cW$-solitons to classical probes by analyzing null geodesics, gravitational lensing, and test scalar perturbations. Across all these tests, $\cW$-solitons closely resemble classical black holes:
\begin{itemize}
    \item They possess a single unstable photon sphere, coinciding with the bubble locus, and admit no stable photon orbits.
    \item Their gravitational lensing displays a photon ring and multiple subrings, qualitatively similar to black hole shadows, albeit with distinctive reflective properties due to the absence of a horizon (which may be reduced once nonlinear tidal disruption near the bubble locus are included~\cite{Heidmann:2025yzd}).
    \item Their scalar QNM spectrum features only short-lived modes, with no long-lived echoes typically associated with ultracompact horizonless objects with stable light rings~\cite{Cardoso:2014sna,Cardoso:2016rao,Cardoso:2017cqb,Cardoso:2019rvt,Heidmann:2023ojf}.
    \item Their linear response in the time domain exhibits a ringdown qualitatively similar to that of the corresponding black hole, with quantitative differences in oscillation frequency and damping rate.
\end{itemize}
We have also provided evidence that $\cW$-solitons and their corresponding black holes are linearly stable under test scalar perturbations.

Beyond these phenomenological similarities, $\cW$-solitons display quantitative departures from black holes in observables such as the photon-sphere radius, Lyapunov exponents, QNM frequencies, and ringdown waveforms. These deviations may serve as potential observational signatures in future high-precision gravitational-wave or electromagnetic measurements, offering novel tests of the black hole paradigm and probes of horizon-scale black hole microstructure.

In this context, a powerful diagnostic of the nature of the remnant formed in a binary merger is provided by gravitational-wave ringdown spectroscopy~\cite{Berti:2025hly}. The LIGO–Virgo–KAGRA Collaboration recently performed the first robust spectroscopy test using the exceptional GW250114 event~\cite{LIGOScientific:2025epi,LIGOScientific:2025obp}, measuring the first overtone with an accuracy of approximately $30\%$. Interestingly, this is only slightly larger than the deviations in the QNMs and in the ringdown waveform predicted for $\cW$-solitons. Although our study focused on test scalar fields and nonspinning solutions, we expect similar differences for gravitational QNMs of spinning solitons. Computing the full ringdown spectrum in that case remains technically challenging but is essential to enable meaningful comparisons with data.
Furthermore, the fundamental QNM of GW250114 has been measured with better accuracy than the first overtone, at the level of $6\%$ and $10\%$, for the real and imaginary part, respectively~\cite{LIGOScientific:2025epi,LIGOScientific:2025obp}. These precise measurements already permit accurate determinations of the remnant’s mass and spin, enabling stringent inspiral–merger–ringdown consistency tests~\cite{LIGOScientific:2025obp}. This highlights the importance of developing similar tests in the present context, where the inspiral and merger dynamics of $\cW$-solitons remain unexplored.
Overall, our analysis indicates that excluding $\cW$-solitons, or alternatively identifying their observational signatures, may be within reach of forthcoming observations.  

Our results open several directions for future research. Extending the analysis to gravitational and electromagnetic perturbations would allow us to capture more realistic features of gravitational-wave signals and assess the complete linear stability of these solutions (see \cite{Guo:2022rms,Cipriani:2024ygw,Bena:2024hoh,Dima:2024cok} for similar analysis for topological stars). It is also natural to investigate their thermodynamic stability, interpreting the $\cW$-soliton and the black string as saddle points of the Euclidean gravitational path integral~\cite{York:1986it,Whiting:1988qr,Braden:1987ad,Bah:2021irr}. Since black holes are thermodynamically unstable (with negative specific heat, as in the Schwarzschild case), one may expect related instabilities for solitons, which merit further exploration.

In this paper, we focused on effectively massless modes from a four-dimensional perspective, which requires the Kaluza-Klein momentum $p$ to vanish. While modes with $p \neq 0$ are not phenomenologically motivated when $R_\psi$ is small---as they correspond to highly energetic perturbations---we have nevertheless found evidence for the existence of bound states in the $\cW$-soliton. A detailed study of their associated normal modes is therefore a worthwhile direction for future work.

The inclusion of rotation remains another crucial challenge, both technically and phenomenologically, given that most astrophysical black holes carry significant angular momentum. Constructing rotating, nonextremal topological solitons in supergravity that remain comparable to black holes has long been difficult, yet recent progress~\cite{Chakraborty:2025ger} suggests that viable rotating generalizations may soon be within reach 
(see~\cite{Bianchi:2025uis} for recent rotating solutions which asymptote to $S^1$ over a magnetic Melvin universe, and \cite{Heidmann:2025pbb} for rotating asymptotically flat solutions).

Finally, $\cW$-solitons provide a fertile arena for studying nonlinear dynamics in horizonless microstate geometries. Their formation, interaction with matter, potential nonlinear instabilities, and implications for information retrieval are promising avenues toward a deeper understanding of the fundamental nature of classical black holes.

\begin{acknowledgements}
We thank Ibrahima Bah, Soumangsu Chakraborty, and David Pereñiguez for valuable discussions.
    A.D., M.M., and P.P. are partially supported by the MUR FIS2 Advanced Grant ET-NOW (CUP:~B53C25001080001) and by the INFN TEONGRAV initiative. P.H. and G.P. are supported by the Department of Physics at The Ohio State University. 
    Some numerical computations have been performed at the Vera cluster supported by the Italian Ministry of Research and by Sapienza University of Rome.
\end{acknowledgements}

\appendix

\section{Spherical coordinates of the $\mathcal{W}$-soliton}
\label{app:SpherCoorWSol}

The coordinates used in \eqref{eq:W_sol_metric} are not genuinely spherical, since the radial variable does not coincide with the radius of the two–sphere. Introducing proper spherical coordinates $(R,\theta,\phi)$ requires the redefinition
\begin{equation}
    R=\sqrt{r(r-2M)}\quad \Leftrightarrow\quad r=\sqrt{R^2+M^2}+M,
\end{equation}
after which the metric and gauge field become
\begin{align}
    ds^2_\cW &= h_+ h_- \left(d\psi+\frac{Q\left(\sqrt{R^2+M^2}-M \right)}{R^2\,h_+} dt+Q(\cos \theta+1) \,d\phi \right)^2 - \frac{dt^2}{h_+}+ \frac{R^2\,dR^2}{(R^2+M^2)h_-}+R^2 \left( d\theta^2 +\sin^2 \theta\,d\phi^2 \right) ,\nonumber \\
    A &= \frac{Q\left(\sqrt{R^2+M^2}-M \right)}{R^2}\,dt+ (1-h_-)\left(d\psi+Q (\cos \theta+1) \,d\phi \right)- Q (\cos \theta+1) \,d\phi\,,
    \label{eq:W_sol_metricApp}
\end{align}
where we have defined:
\begin{equation}
    h_\pm \equiv 1\pm 2\, \frac{M(M+\sqrt{R^2+M^2})-Q^2}{R^2}.
\end{equation}


\bibliographystyle{apsrev4-1}
\bibliography{ref.bib}

\end{document}